\documentclass[useAMS,usenatbib]{mn2e}
\pdfoutput=1
\usepackage{amssymb}
\usepackage{amsmath}
\usepackage{natbib}
\usepackage[pdftex]{graphicx}
\usepackage{subfigure}
\usepackage{booktabs}
\usepackage{tabularx}
\begin{document}
\label{firstpage}
 \title[Precession of giant planets]
{On turbulence driven by axial precession and tidal evolution of the spin--orbit angle of close-in giant planets}
          \author[A. J. Barker]{Adrian J. Barker\thanks{Email address: ajb268@cam.ac.uk} \\
Department of Applied Mathematics and Theoretical Physics, University of Cambridge, Centre for Mathematical Sciences, \\ Wilberforce Road, Cambridge CB3 0WA, UK}
\date{}
\pagerange{\pageref{firstpage}--\pageref{lastpage}} \pubyear{2016}
\maketitle

\begin{abstract}
The spin axis of a rotationally deformed planet is forced to precess about its orbital angular momentum vector, due to the tidal gravity of its host star, if these directions are misaligned. This induces internal fluid motions inside the planet that are subject to a hydrodynamic instability.
We study the turbulent damping of precessional fluid motions, as a result of this instability, in the simplest local computational model 
of a giant planet (or star), with and without a weak internal magnetic field. Our aim is to determine the outcome of this instability, and its importance in driving tidal evolution of the spin--orbit angle in precessing planets (and stars). We find that this instability produces turbulent dissipation that is sufficiently strong that it could drive significant tidal evolution of the spin--orbit angle for hot Jupiters with orbital periods shorter than about 10--18 days. If this mechanism acts in isolation, this evolution would be towards alignment or anti-alignment, depending on the initial angle, but the ultimate evolution (if other tidal mechanisms also contribute) is expected to be towards alignment. The turbulent dissipation is proportional to the cube of the precession frequency, so it leads to much slower damping of stellar spin--orbit angles, implying that this instability is unlikely to drive evolution of the spin--orbit angle in stars (either in planetary or close binary systems). We also find that the instability-driven flow can act as a system-scale dynamo, which may play a role in producing magnetic fields in short-period planets.
\end{abstract}


\section{Introduction}
\label{Introduction}

An isolated gaseous planet is oblate due to its axial rotation. If such a planet is placed into orbit about a star with an arbitrary orientation of its spin axis, the tidal forces from its host star would torque the planet and cause its spin axis to precess about the orbit normal vector. 
In principle, this precession may be observable for some short-period transiting planets through the transit depth variations that it would produce \citep{CarterWinn2010b}. Such an observation would provide important constraints on the oblateness and interior structure of a short-period planet \citep{BarnesFortney2003,CarterWinn2010a,CarterWinn2010b,Correia2014,Zhu2014}. However, precession would not be expected if the planetary spin--orbit angle (obliquity) was either initially, or had time to evolve due to tidal (or other) mechanisms, to 0 or 180 degrees (or 90 degrees, but see \S~\ref{secular}). Indeed, we might expect the planetary spin to be aligned if tidal dissipation had synchronised its spin with its orbit, but in principle the evolution of the spin--orbit angle could occur at a somewhat different rate \citep{Lai2012,Ogilvie2014}. The purpose of this paper is to study tidal evolution of planetary (and stellar) spin--orbit angles when the spin axis is precessing.

In recent years, it has become possible to observe the axial precession of stars in a misaligned binary system (e.g.~\citealt{AlbrechtNature2009,Albrecht2014}). Tentative evidence of axial precession from the variable rotational broadening of stellar spectral lines was presented for DI Herculis as far back as the 1980's \citep{ResiGuinan1989}, and this has been subsequently supported by recent studies \citep{AlbrechtNature2009,PhilRaf2013}. Perhaps the clearest evidence of axial precession in stars is in CV Velorum \citep{Albrecht2014}. These observations motivate theoretical studies of axial precession and tidal spin--orbit alignment in gaseous stars. A fluid body may not precess in the same way as a rigid body (e.g.~\citealt{Kopal1959}), but the response of such a body to precession has remained mostly unexplored. In this context, \cite{PapPringle1982} presented pioneering linear calculations to study the response of a gaseous star to precession.

The related problem of the precession of the Earth's liquid outer core has been studied for over a century, in part because of its potential to drive the geodynamo \citep{Malkus1963,Malkus1968,Loper1975,Tilgner2005,WuRoberts2009}. \cite{Poincare1910} derived an elegant laminar solution of uniform vorticity that describes the response of the spheroidal fluid core to precession imposed at the outer boundary (the mantle). This flow was later shown to be unstable to a parametric instability that excites pairs of inertial waves \citep{Kerswell1993,Kerswell2002}, which we will hereafter refer to as the precessional instability. The nonlinear evolution of this instability has been studied in an idealised Cartesian model (with rigid stress-free top and bottom boundaries), as well as in the more realistic spheroidal geometry. In both cases it has been found to drive turbulence and lead to enhanced dissipation over that of the laminar precessional flow \citep{MasonKerswell2002,LorenzaniTilgner2003,WuRoberts2008}. However, the properties of this turbulent state have not been fully explored, and at present it remains unclear what role the instability of precession-driven flows could play in explaining the geodynamo \citep{LeBars2015}.

Since gaseous planets are primarily (or wholly) fluid, the forces that produce spin precession may induce non-trivial global flows inside the planet. To a first approximation, these will be similar in character to Poincar\'{e}'s flow, and therefore will also be subject to hydrodynamical instabilities. We would expect the resulting turbulent dissipation to drive tidal evolution of the spin--orbit angle to eliminate the precession. This angle would therefore tend to evolve towards 0 or 180 degrees, at least if this mechanism operated in isolation (in reality, other tidal mechanisms will also intervene, so that the eventual evolution will be towards alignment). In this paper we adopt the idealised Cartesian model of \cite{MasonKerswell2002} (without solid boundaries since we are interested in studying giant planets and stars) to study the nonlinear outcome of these instabilities, and to quantify their efficiency at driving tidal spin--orbit evolution. This model can be thought to represent a ``small-patch" within the body of a precessing giant planet or star and allows us to study the properties of the turbulent flow driven by the precessional instability. A similar model has recently been used to study the related elliptical instability \citep{BL2013,BL2014}.

Short-period gaseous planets are very likely to generate their own internal magnetic fields, just like the giant planets in our Solar system \citep{Jones2011}. Currently, there is no conclusive observational evidence of magnetic fields in extrasolar planets, but their external manifestations might be detectable through radio emission caused by star-planet magnetic interaction \citep{Stevens2005,Zarka2007}, or through causing asymmetries during transits (e.g.~\citealt{Vidotto2010}). Recently, there have been tentative indications of magnetic fields in short-period extrasolar planets from absorption features during transits \citep{Vidotto2010,Kislyakova2014,Cauley2015} which suggest these planets to have dipolar fields of strength $B\sim 20-30$ G, which is roughly similar to the observed dipolar field at the surface of Jupiter. Given that short-period gaseous planets are very likely to be magnetised, we will study the nonlinear outcome of the precessional instability in the presence of a magnetic field. A secondary aim of this work is to investigate whether the precessional instability may play a role in generating the magnetic fields of giant planets. Indeed, there are indications that the related elliptical instability could drive a dynamo \citep{BL2014,Cebron2014}. 

We first outline how planetary spins are thought to secularly precess due to stellar tidal forces in \S~\ref{secular}, before presenting simple estimates of the turbulent dissipation resulting from the precessional instability in \S~\ref{turb}, where we also estimate its ability to produce tidal evolution of the spin--orbit angle. We describe our idealised model and our tests of its implementation in \S~\ref{localmodel}, before presenting the results of our simulations in \S~\ref{results} and \S~\ref{synthesis}. Fig.~\ref{4} is our main result, where we compare the results of our simulations against the simple predictions in \S~\ref{turb}. We finish with a discussion and conclusion.

\section{Planetary spins precess due to stellar tidal torques}
\label{secular}
We consider a gaseous planet of mass $M_p$, and radius $R_p$, which rotates axially with uniform angular velocity $\boldsymbol{\Omega}=\Omega\boldsymbol{\hat{\Omega}}$, and orbits a star of mass $M_\star$. Its orbit has eccentricity $e$, and specific orbital angular momentum $\boldsymbol{h}=h\hat{\boldsymbol{h}}$, such that the planetary spin--orbit angle (obliquity) is $\psi=\cos^{-1}(\hat{\boldsymbol{h}}\cdot \hat{\boldsymbol{\Omega}})$. We define the planetary spin angular momentum to be $I_3 \boldsymbol{\Omega}$, where $I_3=r_g^2 M_p R_p^2$, and $r_g^2 \approx 0.26$ is the squared radius of gyration (of a polytrope of index 1; this is consistent with the inferred value for Jupiter, e.g.~\citealt{Helled2011}). The tidal torque due to the star will cause the planetary spin axis to precess about the total angular momentum vector $\boldsymbol{L}=\mu\boldsymbol{h}+I_3\boldsymbol{\Omega}\approx \mu\boldsymbol{h}$ (since $I_3\Omega\ll \mu h$, where the reduced mass is $\mu=\frac{M_p M_{\star}}{M_p+M_\star}$), unless $\boldsymbol{L}$ is initially perfectly aligned, anti-aligned, or perpendicular to $\hat{\boldsymbol{\Omega}}$, i.e. if $\cos\psi \ne \left\{0,\pm 1\right\}$. If we average this (non-dissipative) tidal torque over an orbit, we obtain the secular evolutionary equation
\begin{eqnarray}
\frac{\mathrm{d}\boldsymbol{\hat{\Omega}}}{\mathrm{d}t} = -\frac{\mu h}{I_3\Omega}\frac{\mathrm{d}\boldsymbol{\hat{h}}}{\mathrm{d}t}
=-\Omega_p \boldsymbol{\hat{h}}\times\boldsymbol{\hat{\Omega}},
\end{eqnarray}
where the precession frequency is \citep{Kopal1959,Eggleton2001,Goldstein2002}
\begin{eqnarray}
\Omega_p &=& \frac{3 G M_\star}{2a^3\Omega}\frac{(I_3-I_1)}{I_3}\frac{\cos\psi}{(1-e^2)^{\frac{3}{2}}}.
\end{eqnarray}
Note that the magnitudes $\Omega$ and $h$ are preserved in the absence of dissipation. In a similar manner, $\hat{\boldsymbol{h}}$ precesses around $\boldsymbol{L}$ but at the much slower rate $\frac{I_3\Omega}{\mu h}\Omega_p$.
We define $n=\frac{2\pi}{P}=\sqrt{\frac{G(M_{\star}+M_p)}{a^3}}$, to be the orbital mean motion, $P$ to be the orbital period, and $a$ to be the semi-major axis. We also define the dimensionless tidal amplitude
\begin{eqnarray}
\epsilon_T=\left(\frac{M_{\star}}{M_p}\right)\left(\frac{R_p}{a}\right)^3=\frac{M_\star}{M_p+M_\star}\left(\frac{P_\mathrm{dyn}}{P}\right)^2,
\end{eqnarray}
where $P_{\mathrm{dyn}}=\frac{2\pi}{\omega_\mathrm{dyn}}$, and the dynamical frequency is $\omega_\mathrm{dyn}=\sqrt{\frac{GM_p}{R_p^3}}$.

Faster rotation makes gaseous bodies more oblate, such that
\begin{eqnarray}
I_3-I_1 \approx K M_p R_p^2 \left(\frac{\Omega}{\omega_\mathrm{dyn}}\right)^2,
\end{eqnarray}
 where $K=\frac{k_2}{3}=J_2 \frac{\omega_\mathrm{dyn}^2}{\Omega^2}$, $k_2$ is the second-order Love number and $J_2$ is the second harmonic coefficient. It is observed that $J_2\approx0.015$ (which is approximately consistent with $k_2\approx0.38$) for Jupiter \citep{Guillot2004}, so that $K\approx 0.125$. The precession period $P_p$ of a Jupiter-mass planet with spin period $P_\Omega$ that orbits a Sun-like star is
\begin{eqnarray}
\label{precperiod}
P_p &\equiv& \frac{2\pi}{\Omega_p} = \frac{2r_g^2}{3K}\frac{P_\Omega}{\epsilon_T} \frac{\left(1-e^2\right)^{\frac{3}{2}}}{\cos\psi}, \\
&\approx& 1.3\,\mathrm{yr} \left(\frac{P}{10\,\mathrm{d}}\right)^2\left(\frac{P_\Omega}{10 \,\mathrm{hr}}\right) \frac{\left(1-e^2\right)^{\frac{3}{2}}}{\cos\psi}.
\label{precperiod2}
\end{eqnarray}
This precession is very slow in comparison with the spin of the planet, but the spin vector will precess many times around the orbital angular momentum vector during the main-sequence lifetime of the star, at least until the spin--orbit angle has evolved to satisfy $\cos\psi=\{0,\pm1\}$. The question that we wish to address is: how long would it take for the spin--orbit angle to undergo significant tidal evolution due to the precessional instability?

\section{Precession-driven turbulence and tidal spin--orbit evolution: simple estimates}
\label{turb}

A uniformly precessing flow is known to be unstable \citep{Kerswell1993,Kerswell2002}. This is because the precessional flow in the fluid frame is time-periodic (with frequency of magnitude $\Omega$), and this periodic variation can excite pairs of inertial waves (with frequencies $\omega_1$ and $\omega_2$, such that $|\omega_1\pm\omega_2|\approx\Omega$) in parametric resonance. The fastest growing modes typically have $|\omega_1|\approx|\omega_2|\approx\frac{\Omega}{2}$ (at least for short wavelength modes, but the fastest growing global modes may have somewhat different frequencies e.g.~\citealt{Lin2015}), and their growth rates are
\begin{eqnarray}
\sigma \sim \Omega_p = \epsilon \Omega.
\end{eqnarray}
Since the planet precesses very slowly, we define (the ``Poincare number") $\epsilon\equiv\frac{\Omega_p}{\Omega}\ll 1$ ($\epsilon\approx9\times 10^{-4}$, according to Eq.~\ref{precperiod2}). If such an instability grows until the unstable mode amplitudes are limited by their own shear instabilities, and we obtain a statistically steady turbulent cascade, this will have
$\sigma \sim \frac{u}{\lambda}$,
where $u$ and $\lambda$ are a typical velocity and lengthscale for the energetically-dominant (``outer scale") modes, respectively. This implies
\begin{eqnarray}
\label{turbvel}
u\sim \epsilon \Omega \lambda.
\end{eqnarray}
The corresponding viscous dissipation rate is
\begin{eqnarray}
D\sim M_p \frac{u^3}{\lambda} \sim M_p \epsilon^3\Omega^3\lambda^2.
\end{eqnarray}
We define an efficiency factor $\chi$, such that
\begin{eqnarray}
\label{disspred}
D\equiv \chi M_p\epsilon^3\Omega^3 R_p^2\sim \chi \rho R_p^5\Omega_p^3,
\end{eqnarray}
which quantifies the efficiency of the turbulent dissipation\footnote{This is much smaller than the rigorous upper bound derived by \cite{Kerswell1996} of approximately $0.43 \rho R_p^5 \Omega^3$, which must be inapplicable to the \textit{bulk} dissipation, since it is independent of $\Omega_p$.}, and is most efficient if $\chi\sim 1$ (or larger, if this is possible). It is the primary aim of this paper to determine whether these scaling laws adequately describe the turbulence driven by the precessional instability. For our simulations, we define $D\equiv \chi \epsilon^3$ and $u\equiv C\epsilon$, where $\chi$ and $C$ (and their possible dependences on $\epsilon$) are to be determined numerically. We will later (in Fig.~\ref{4}) present evidence for the validity of these scalings (with $\chi$ and $C$ being approximately constant) over a range of $\epsilon\in [0.01,0.5]$ that can be probed numerically.

The dissipation is associated with a tidal torque that will drive evolution of the planetary spin--orbit angle. To obtain a crude estimate of the efficiency of this damping process, we derive a tidal quality factor for this component in the case of a circular orbit (the magnitude of the relevant tidal frequency is $\Omega$)
\begin{eqnarray}
Q&=&
\frac{\Omega E_0}{D} 
= \frac{\epsilon_T^2}{\chi}\frac{P_p^3}{P_\mathrm{dyn}^2P_\Omega},
\end{eqnarray}
where the maximum energy stored in the tidal distortion is $E_0\approx\frac{GM_p^2}{R_p}\epsilon_T^2$.
For this component, the spin--orbit angle evolves according to \citep{Lai2012,Ogilvie2014}, 
\begin{eqnarray}
\frac{\mathrm{d}\psi}{\mathrm{d} t}=-\frac{\sin\psi \cos^2 \psi}{\tau_{10}}\left(\cos\psi+\frac{I_3\Omega}{\mu h}\right),
\end{eqnarray}
where
\begin{eqnarray}
\frac{1}{\tau_{10}}&=&\frac{3}{4}\frac{k_2}{Q}\left(\frac{M_{\star}}{M_p}\right)\left(\frac{R_p}{a}\right)^5 \left(\frac{\mu h}{I_3\Omega}\right) n \\
&=& \frac{3\pi}{2r_g^2}\frac{k_2}{Q}\epsilon^2_T \frac{P_\Omega}{P_\mathrm{dyn}^2}.
\end{eqnarray}
For small angles, we obtain exponential damping of the misalignment on a timescale
\begin{eqnarray}
\tau_{\psi}&=&\tau_{10}\left(1+\frac{I_3\Omega}{\mu h}\right)^{-1} \\
&=&\frac{2}{3\pi\chi}\frac{r_g^2}{k_2}\frac{P_p^3}{P_{\Omega}^2}\left(1+\frac{I_3\Omega}{\mu h}\right)^{-1} \\
&=& \frac{16 r_g^8}{81\pi \chi k_2 K^3} \frac{P_\Omega}{\epsilon_T^3}\left(1+\frac{I_3\Omega}{\mu h}\right)^{-1} \\
&\approx& 10^{9}\,\mathrm{yr}\,\left(\frac{10^{-2}}{\chi}\right)\left(\frac{P}{18\,\mathrm{d}}\right)^6\left(\frac{P_{\Omega}}{10\,\mathrm{hr}}\right),
\label{taupred}
\end{eqnarray}
for a typical hot Jupiter around a Sun-like star\footnote{This can be compared with the timescale that that we have previously estimated for the related elliptical instability \citep{Barker2015b}. In that case, simulations suggest the alignment timescale to be comparable with the synchronisation timescale, so that
\begin{eqnarray}
\tau_{\psi}\sim\tau_\Omega\approx \frac{r_g^2}{18\pi \chi_E} \frac{P_\Omega}{\epsilon_T^3},
\end{eqnarray}
where $\chi_E\approx 10^{-2}-10^{-1}$, is the equivalent dissipation efficiency. This has the same functional dependence as the timescale due to the precessional instability because $\Omega_p\propto \epsilon_T\Omega$, so that the growth rates for both instabilities are proportional to $\epsilon_T\Omega$. In fact, the elliptical instability would similarly predict alignment out to approximately 15 days. The precessional instability would operate in addition to the elliptical instability if the planetary spin axis is initially misaligned.}.
The value chosen for $\chi$ is approximately what is suggested by the simulations that we will present in \S~\ref{results} (Fig.~\ref{4}). 
On this timescale, we would expect the planetary spin to become aligned with its orbit. For nearly anti-aligned orbits, the evolution towards anti-alignment would occur on the same timescale. For nearly perpendicular\footnote{Due to this mechanism acting in isolation, $\psi=90^{\circ}$ is an unstable equilibrium (whereas $\psi=0^{\circ}$ or $180^{\circ}$ are stable equilibria if $I_3\Omega\ll \mu h$), so that an orbit that is nearly perpendicular evolves towards $\psi=0^{\circ}$ or $180^{\circ}$ and not towards $90^{\circ}$ (cf.~\citealt{RogersLin2013}; see also \citealt{LiWinn2016}). However, the evolutionary timescale if $\psi\approx90^{\circ}$ initially is very long (relative to that for nearly aligned or anti-aligned cases), so a random distribution of initial $\psi$ would be expected to show clustering around $90^{\circ}$ in addition to $0^{\circ}$ and $180^{\circ}$ \citep{Lai2012,RogersLin2013}.} orbits, the spin--orbit evolution towards alignment or anti-alignment occurs on the much longer timescale $\tau_{10}\frac{\mu h}{I_3\Omega}$. This estimate indicates that the hydrodynamic instability of the precessional flow inside the planet could be important in driving evolution of the planetary spin--orbit angle for observed hot Jupiters.

On the other hand, this instability in the star is unlikely to play a role in the evolution of the stellar spin--orbit angle with a Jupiter-mass planetary companion, since in that case we obtain
\begin{eqnarray}
\tau_{\psi}\approx 4\times10^{17}\,\mathrm{yr}\,\left(\frac{10^{-2}}{\chi}\right)\left(\frac{P}{1\,\mathrm{d}}\right)^6\left(\frac{P_{\Omega}}{10\,\mathrm{d}}\right).
\end{eqnarray}
for a solar-type star with $r_g^2\approx 0.1$, and $K\approx0.05$ \citep{ClaretGimenez1992,StorchAndersonLai2014}. In a close binary system, we also expect the stellar spin--orbit evolution to be negligible based on a similar estimate.

\subsection{Presence of magnetic fields}

Before we begin to describe our numerical setup, we will now crudely estimate the relative strength of planetary magnetic fields, to determine whether they may be important. A typical Alfv\'{e}n speed for the planetary magnetic field $v_A\sim \frac{B}{\sqrt{\mu_0\bar{\rho}}}$ (where $\bar{\rho}$ is the mean density), can be compared with the expected turbulent velocity due to precession-instability-driven flows (Eq.~\ref{turbvel}) to obtain
\begin{eqnarray}
\frac{v_A}{u}&\approx& \frac{B}{\Omega_p R\sqrt{\mu_0 \bar{\rho}}} \\
&\approx& 2\times10^{-3}\left(\frac{B}{10\,\mathrm{G}}\right)\left(\frac{1\mathrm{g cm^{-3}}}{\bar{\rho}}\right)\left(\frac{P}{10\,\mathrm{d}}\right)^2\left(\frac{P_{\Omega}}{10\,\mathrm{hr}}\right),
\end{eqnarray}
for Jupiter in a short-period orbit (taking a magnetic field strength consistent with Jupiter's dipolar magnetic field at the surface). Magnetic fields are therefore expected to be weak in comparison with the precessionally-driven flows. However, even a weak magnetic field can drastically alter the properties of the turbulence, as we will demonstrate below. 

For our MHD simulations, we choose an initial magnetic field strength $B_0\ll \epsilon$, so that we are in the weak-field regime. Note that in the case of stars hosting short-period planets, it may be that $v_A\gtrsim u$, therefore we would no longer be in the weak field regime presented here (in this case the precessional instability can be modified by the magnetic field -- see e.g.~\citealt{Salhi2010}). If Lorentz forces play a non-negligible role in the turbulent state, we might expect $|\boldsymbol{B}\cdot \nabla \boldsymbol{B}|\sim |\boldsymbol{u}\cdot \nabla \boldsymbol{u}|$, which suggests a scaling like $B\sim u\sim \epsilon \Omega \lambda$ (in our simulations below we define $B\equiv C_B \epsilon$, where $C_B$ is to be determined numerically). It is a secondary aim of this paper to determine whether the precessional instability can generate magnetic fields, and whether such a scaling law may adequately capture its ability to drive a dynamo. 

\section{Local model of precession}
\label{localmodel}
We consider a Cartesian model that can be thought to represent a small patch in a convective region of a gaseous planet (or star). The planet is assumed to be uniformly rotating, with angular velocity $\Omega\boldsymbol{e}_{\tilde{z}}$, which is subject to slow ($\epsilon\ll 1$), steady, and uniform precession with angular velocity\footnote{The component of precession along $\boldsymbol{e}_{\tilde{z}}$ does not drive these instabilities, so we neglect it \citep{Kerswell1993}.} $\epsilon \Omega \boldsymbol{e}_{\tilde{x}}$. In the precessing frame with coordinates $(\tilde{x},\tilde{y},\tilde{z})$, the equations of motion for an inviscid, neutrally stratified, incompressible fluid are (we assume the density, $\rho\equiv 1$)
\begin{eqnarray}
&& \partial_t \boldsymbol{v} + \boldsymbol{v}\cdot \nabla \boldsymbol{v} + 2 \epsilon \Omega \boldsymbol{e}_{\tilde{x}} \times \boldsymbol{v} = -\nabla p, \\
&& \nabla \cdot \boldsymbol{v}=0,
\end{eqnarray}
where $\boldsymbol{v}$ is the fluid velocity and $p$ is a modified pressure. The assumption of incompressibility is appropriate (at least in the local model) because the precessional instability excites inertial waves, and the assumption of neutral stratification is approximately valid if convection is efficient. We relegate studying the precessional instability in the presence of turbulent convection (or stable stratification) to future work. We also neglect the tidal deformation of the streamlines in the planet (so that they are circles) in order to isolate the effects of precession.

The precessional flow which is a nonlinear solution of these equations is \citep{Kerswell1993,Kerswell2002}
\begin{eqnarray}
\boldsymbol{V}_{0} = \Omega\left(\begin{array}{lcr}
\label{basicflowold}
0 & -1 & 0 \\
1 & 0 & -2\epsilon \\
0 & 0 & 0 \\
\end{array}\right)\tilde{\boldsymbol{x}}.
\end{eqnarray}
Note that the precession has induced a vertical shear $-2\epsilon \Omega \tilde{z}\boldsymbol{e}_{\tilde{y}}$. It is this shear that drives a hydrodynamic instability. Eq.~\ref{basicflowold} approximates the laminar Poincar\'{e} flow far from the boundaries of a spheroid if $\epsilon\ll 1$, and the tidal deformation is neglected \citep{Poincare1910,Kerswell2002,SalhiCambon2009}.

From now on we take $\Omega=1$ and $L=1$ (the size of our Cartesian box) to define our units. We also find it convenient to transform to the frame in which the mean total angular velocity of the fluid is along the new direction $\boldsymbol{e}_z$, with coordinates $(x,y,z)$ (this is the ``mantle frame" of the precessing planet, using the analogy of the Earth's core, where $x=\tilde{x}\cos t+ \tilde{y}\sin t$, $y=-\tilde{x}\sin t+\tilde{y}\cos t$, and $z=\tilde{z}$). In this frame, $\boldsymbol{V}_0$ is transformed into the oscillatory strain flow \citep{MasonKerswell2002}
\begin{eqnarray}
\label{basicflow}
\boldsymbol{U}_0= -2\epsilon\left(\begin{array}{lcr}
0 & 0 & \sin t \\
0 & 0 & \cos t \\
0 & 0 & 0 \\
\end{array}\right)\boldsymbol{x}=\mathrm{A}\boldsymbol{x}.
\end{eqnarray}
In the presence of uniform kinematic viscosity $\nu$ (which may be thought to represent a turbulent viscosity due to convection), this flow has nonzero viscous dissipation $D_\mathrm{lam}=\frac{2\nu}{V}\int e_{ij}e_{ij} \mathrm{d}V=4\nu\epsilon^2$, where $e_{ij}=\frac{1}{2}\left(\partial_{i}U_{0,j}+\partial_j U_{0,i}\right)$. The momentum equation in this frame is
\begin{eqnarray}
\partial_t \boldsymbol{u}_t + \boldsymbol{u}_t\cdot \nabla \boldsymbol{u}_t + 2 \left(\boldsymbol{e}_z+\boldsymbol{\epsilon}(t)\right) \times \boldsymbol{u}_t = -\nabla p+2 z\boldsymbol{\epsilon}(t),
\end{eqnarray}
where $\boldsymbol{u}_t$ is the total angular velocity in this frame and $\boldsymbol{\epsilon}(t)=\epsilon \left(\cos t, 0, \sin t\right)^{T}$.
We seek perturbations with velocity $\boldsymbol{u}$, to this background precessional flow, such that $\boldsymbol{u}_t=\boldsymbol{U}_0+\boldsymbol{u}$. These perturbations satisfy
\begin{eqnarray}
\nonumber
\partial_t \boldsymbol{u}+ \boldsymbol{u}\cdot \nabla \boldsymbol{u} + 2 \boldsymbol{e}_z\times \boldsymbol{u} +\nabla p&=& -\mathrm{A}\boldsymbol{u}-2\boldsymbol{\epsilon}(t)\times\boldsymbol{u}\\
&&-\mathrm{A}\boldsymbol{x}\cdot\nabla \boldsymbol{u},
\label{perturbationeqns}
\end{eqnarray} 
where all terms related to the precession appear on the RHS. In our simulations below, we also add explicit viscosity (if $\alpha=1$, otherwise this is a ``hyper-viscosity") to the RHS of Eq.~\ref{perturbationeqns}, of the form $\nu_\alpha (-1)^{\alpha+1}\nabla^{2\alpha}\boldsymbol{u}$. The extension of this model to magnetohydrodynamics (as well as the inclusion of ohmic diffusivity or hyperdiffusion), is presented in Appendix~\ref{MHD}. 

We wish to solve Eq.~\ref{perturbationeqns} numerically in a periodic domain that might represent a small-patch of a giant planet (or star). However, the final term is linear in $z$, so we cannot directly apply periodic boundary conditions. We can overcome this problem by decomposing the flow into time-dependent ``shearing waves", which is equivalent to applying periodic boundary conditions in the frame that co-moves with the flow $\boldsymbol{U}_0$. This is similar to the approach used for the ``shearing box" to describe the local dynamics of astrophysical discs \citep{GoldreichLyndenBell1965,Hawley1995}.

\subsection{Spectral decomposition}
\label{spectral}
To eliminate the final term in Eq.~\ref{perturbationeqns}, we seek solutions with time-dependent wavevectors $\boldsymbol{u}=\mathrm{Re}\left[\hat{\boldsymbol{u}}\, \mathrm{e}^{\mathrm{i}\boldsymbol{k}(t)\cdot \boldsymbol{x}}\right]$ (and similarly for $p$, where hats denote Fourier transforms), that satisfy
\begin{eqnarray}
&& \partial_t \hat{\boldsymbol{u}} +\widehat{\boldsymbol{u}\cdot \nabla \boldsymbol{u}} + 2 \boldsymbol{e}_z\times \hat{\boldsymbol{u}} =
-\mathrm{i}\boldsymbol{k} \hat{p} - \mathrm{A}\hat{\boldsymbol{u}}-2\boldsymbol{\epsilon}(t)\times\hat{\boldsymbol{u}}, \\
&& \boldsymbol{k}\cdot \hat{\boldsymbol{u}}, \\
&& \partial_t \boldsymbol{k} = -\mathrm{A}^T\boldsymbol{k}.
\end{eqnarray}
The latter implies that
\begin{eqnarray}
\boldsymbol{k}(t)=\left(k_{x,0},k_{y,0},k_{z,0}+2\epsilon\left(-k_{x,0}\cos t+k_{y,0}\sin t \right)\right)^T,
\end{eqnarray}
where $\langle\boldsymbol{k}(t)\rangle=(k_{x,0},k_{y,0},k_{z,0})^T$ is the time-averaged wavevector and $\boldsymbol{k}(t)$ oscillates in time about this vector. We have modified the Fourier spectral code Snoopy \citep{Lesur2005} to use a basis of these time-evolving wavevectors, so that the corresponding shearing periodic boundary conditions are automatically satisfied (similar to \citealt{BL2013,BL2014}). This allows us to solve Eq.~\ref{perturbationeqns} numerically.

\subsection{Parametric instability of precessional flow: test of code}
\begin{figure}
  \begin{center}
    \subfigure{\includegraphics[trim=7cm 1cm 7cm 1cm, clip=true,width=0.35\textwidth]{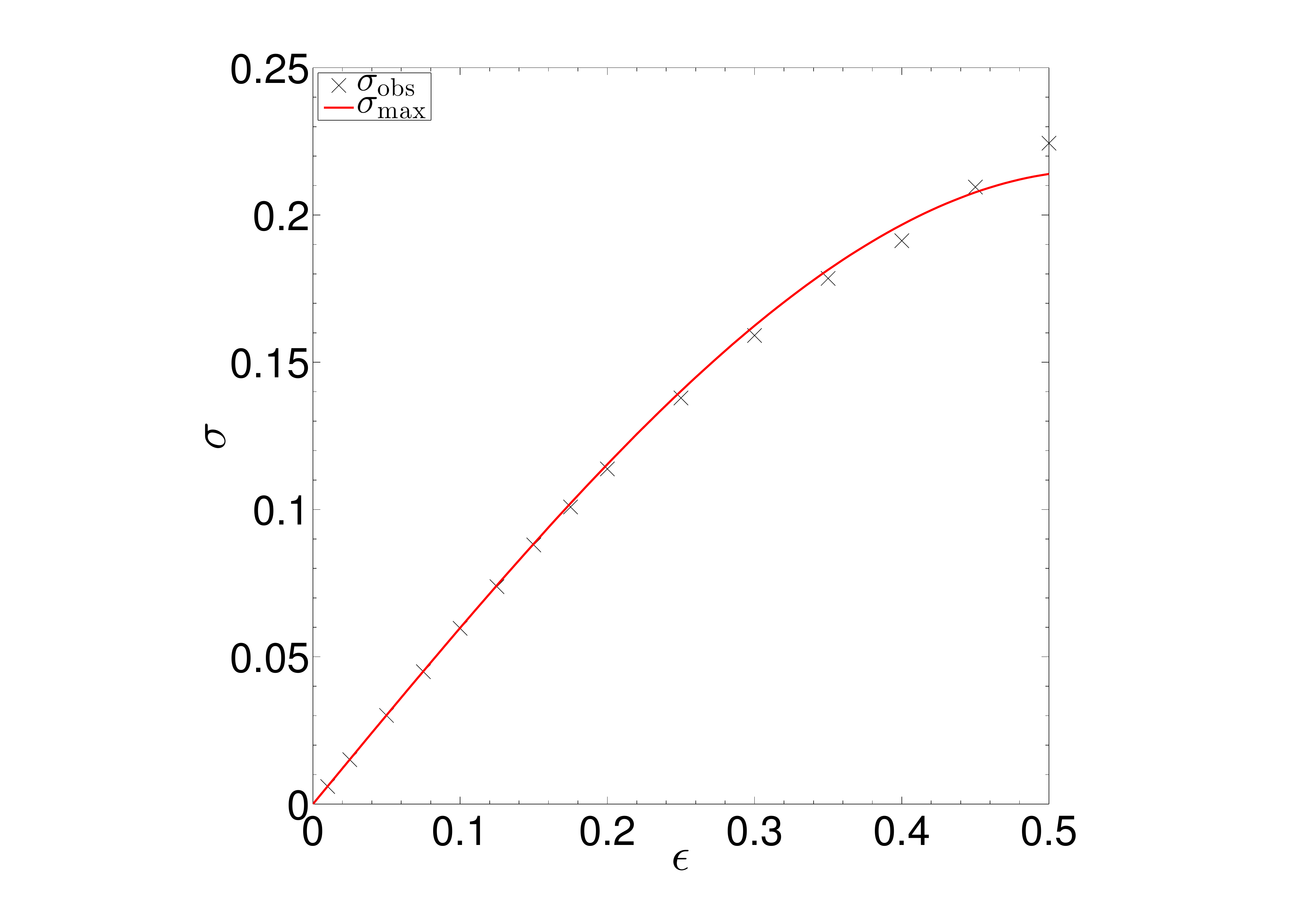}}
    \end{center}
  \caption{Growth rate ($\sigma$; based on half the slope of the linear fit to $\log_{10}K$ as a function of time, where $K=\frac{1}{2}\langle |\boldsymbol{u}|^2\rangle_V$, and $\langle \cdot \rangle_V$ represents a volume-average) as a function of $\epsilon$ from simulations (black crosses), compared with the theoretical prediction ($\sigma_\mathrm{max}$; solid red line). The agreement is excellent for small $\epsilon$, but there are small departures for larger $\epsilon$ which are probably due to the inapplicability of the expansion given by Eq.~\ref{maxgrowth}. All simulations had $64^3$ points, with nonlinear terms switched off, zero viscosity, and initial mode amplitude $10^{-4}$.}
  \label{1}
\end{figure}

The flow given by Eq.~\ref{basicflow} (and Eq.~\ref{basicflowold}) is unstable to the excitation of pairs of inertial waves in a parametric resonance. In this case, the base flow has frequency 1, so the fastest growing subharmonic instabilities involve waves with frequencies $\pm\frac{1}{2}$. Their maximum growth rate for small $|\epsilon|$ is \citep{Kerswell1993,Kerswell2002,NaingFukumoto2011}
\begin{eqnarray}
\label{maxgrowth}
\sigma_\mathrm{max}\approx\frac{5\sqrt{15}}{32}|\epsilon|\left(1-\frac{75}{64}\epsilon^2\right).
\end{eqnarray}
Fig.~\ref{1} shows the growth rate in a set of simulations that have been initialised with the most unstable mode (since our initial $\boldsymbol{k}$ points in the time-averaged direction, we simply have to choose waves that initially satisfy $|\frac{k_z}{k}|\approx\frac{1}{4}$). This shows excellent agreement with the theoretical prediction for small $\epsilon$ (and reasonable agreement for moderate $\epsilon$). We have also confirmed, from a Lomb-Scargle periodogram analysis of the individual velocity components at several test points in the domain, that the magnitude of the frequency in the flow is $\frac{1}{2}$. These provide important tests of the code, so that we can be confident in its application to the nonlinear simulations that we will describe below. 

\section{Nonlinear evolution of the precessional instability}
\label{results}

We initialise the velocity field in our simulations with solenoidal random noise (with amplitude $10^{-4}$ for hyperdiffusive cases with $\alpha\ne 1$, and $10^{-2}$ for those with standard viscosity with $\alpha=1$) for all wavenumbers $\frac{k}{2\pi}\in[1,21]$. If present, the magnetic field is initialised as described in Appendix~\ref{MHD}. A table of simulations is presented in Appendix~\ref{TableSims}. 

\begin{figure}
 \begin{center}
 \subfigure{\includegraphics[trim=0cm 0cm 0cm 0cm, clip=true,width=0.3\textwidth]{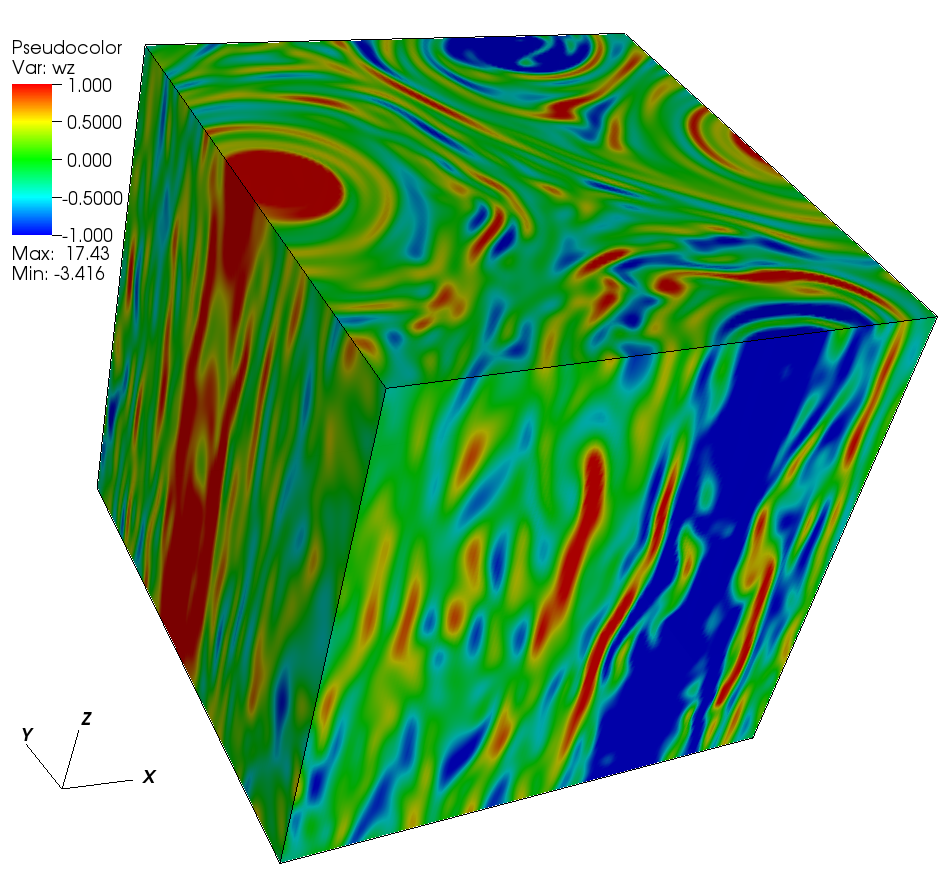}}
  \subfigure{\includegraphics[trim=0cm 0cm 0cm 0cm, clip=true,width=0.3\textwidth]{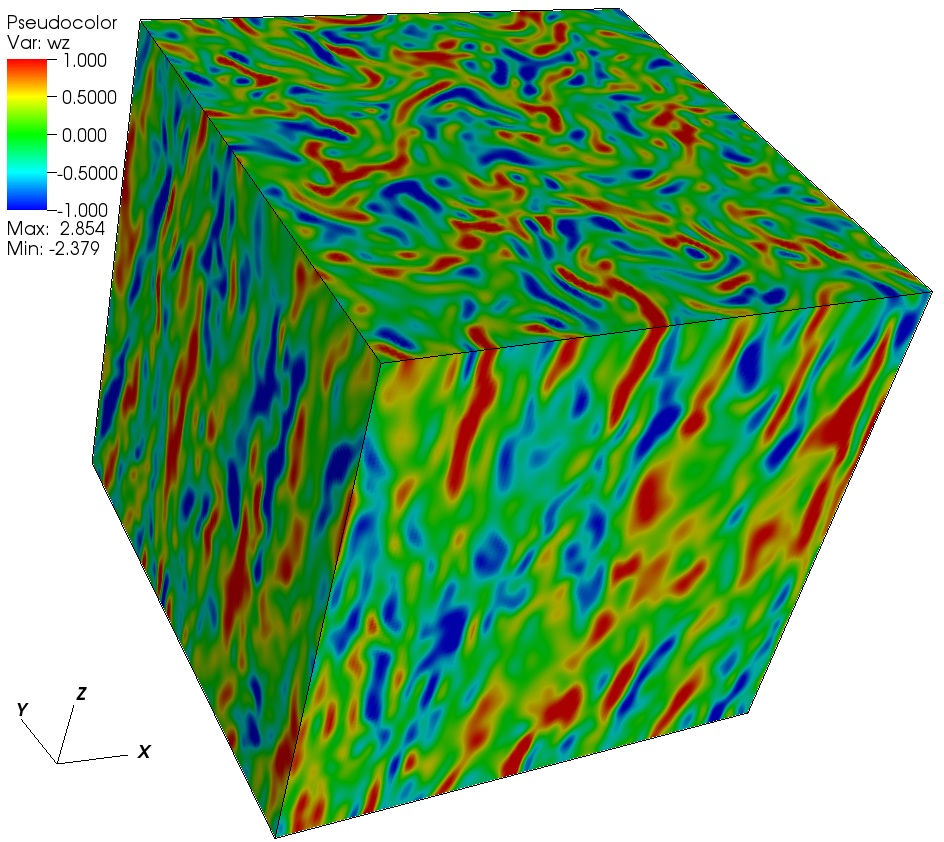}}
 \end{center}
  \caption{Illustration of the vertical vorticity in two simulations with $\epsilon=0.1$, adopting $\alpha=4$ hyperdiffusion and $\nu_{4}=10^{-18}$. Top: hydrodynamical simulation at $t=1100$. Bottom: equivalent MHD simulation at $t=5900$ (with $\eta_{4}=\nu_{4}$). This shows that columnar vortices form and persist in the hydrodynamical simulations, but magnetic stresses prevent their formation in the MHD simulations.}
  \label{3}
\end{figure}

\begin{figure}
 \begin{center}
   \subfigure{\includegraphics[trim=6cm 0cm 7cm 1cm, clip=true,width=0.3\textwidth]{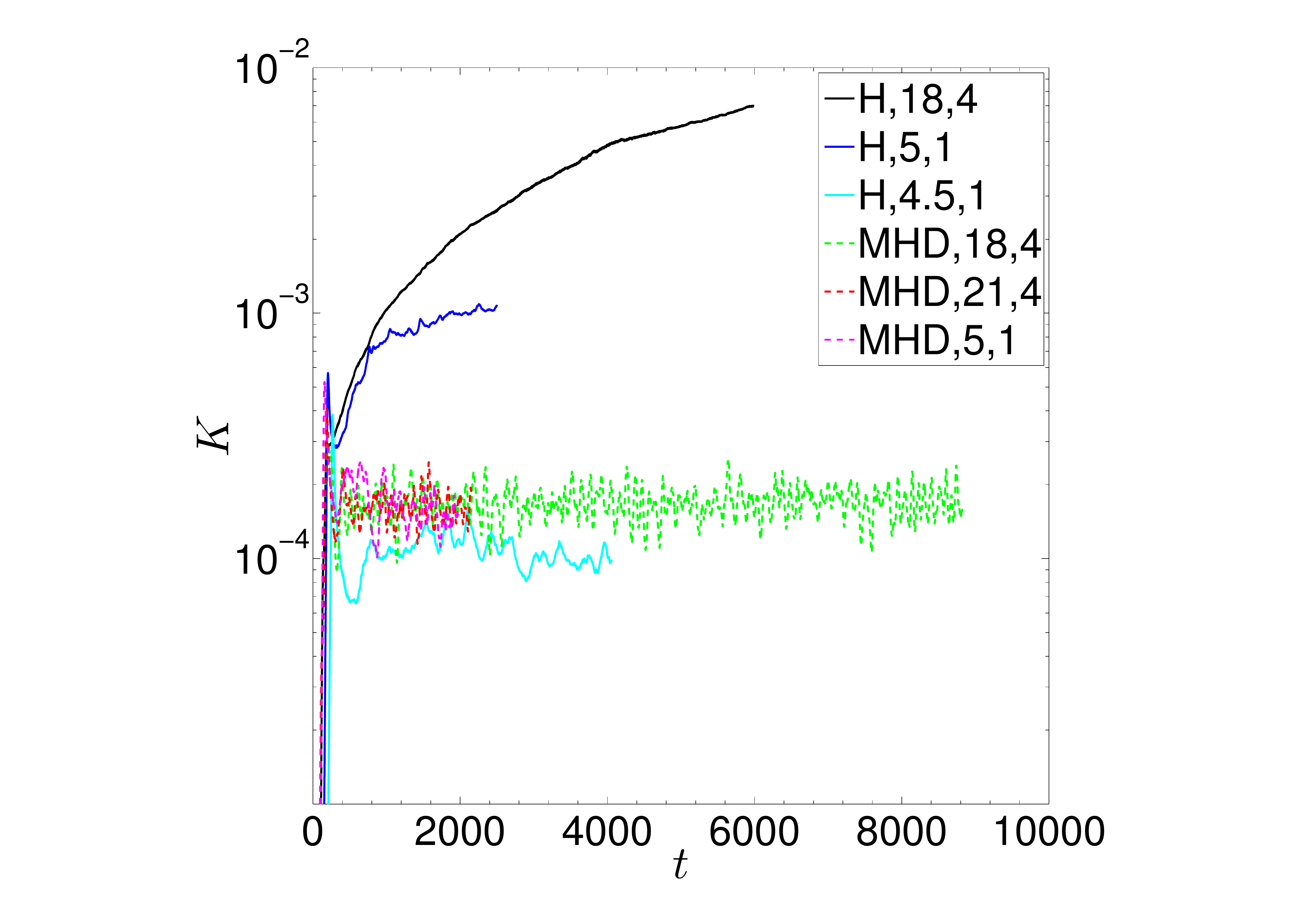}}
   \subfigure{\includegraphics[trim=8cm 0cm 2cm 3cm, clip=true,width=0.3\textwidth]{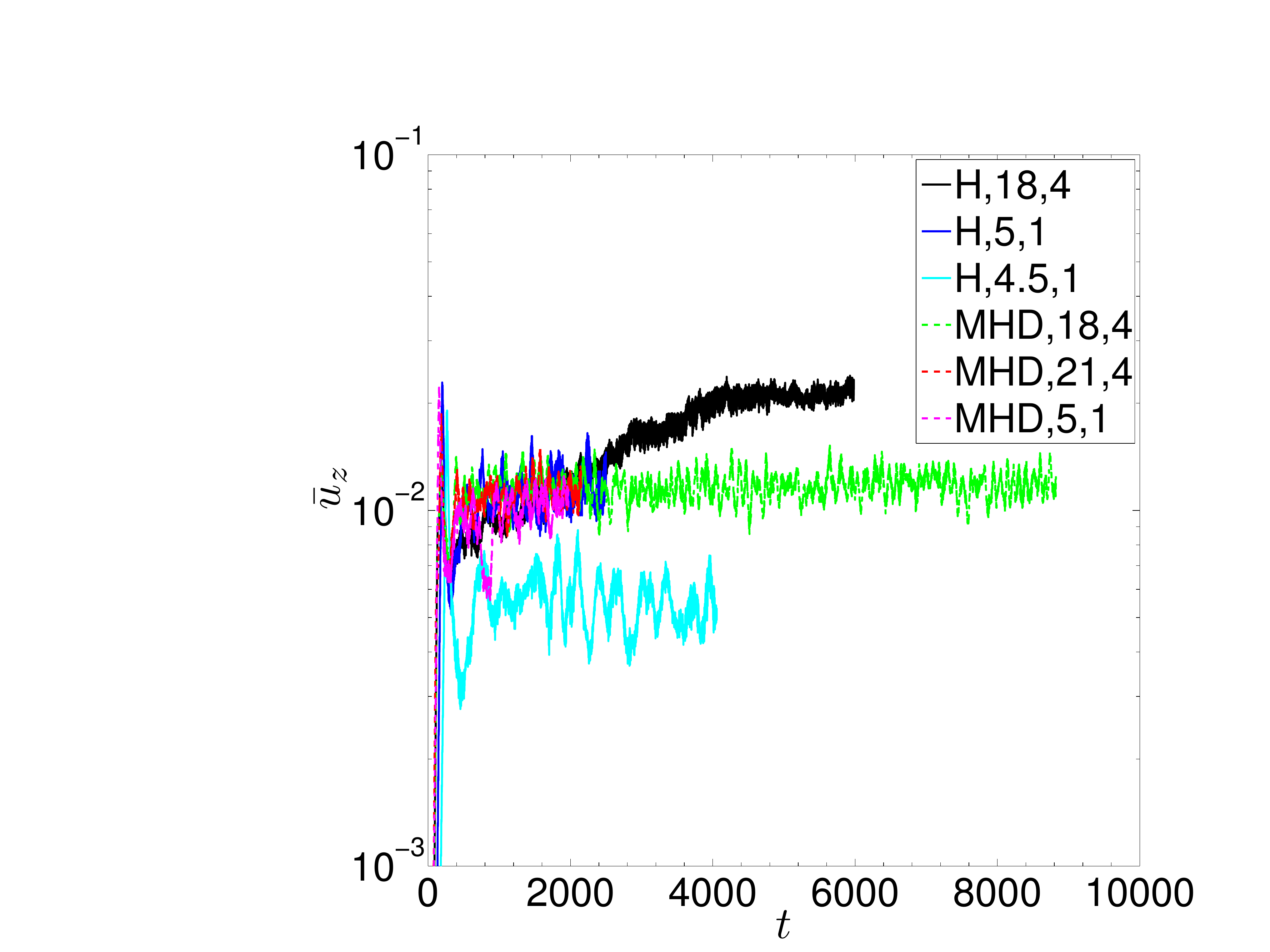}}
   \subfigure{\includegraphics[trim=7cm 0cm 3cm 3cm, clip=true,width=0.3\textwidth]{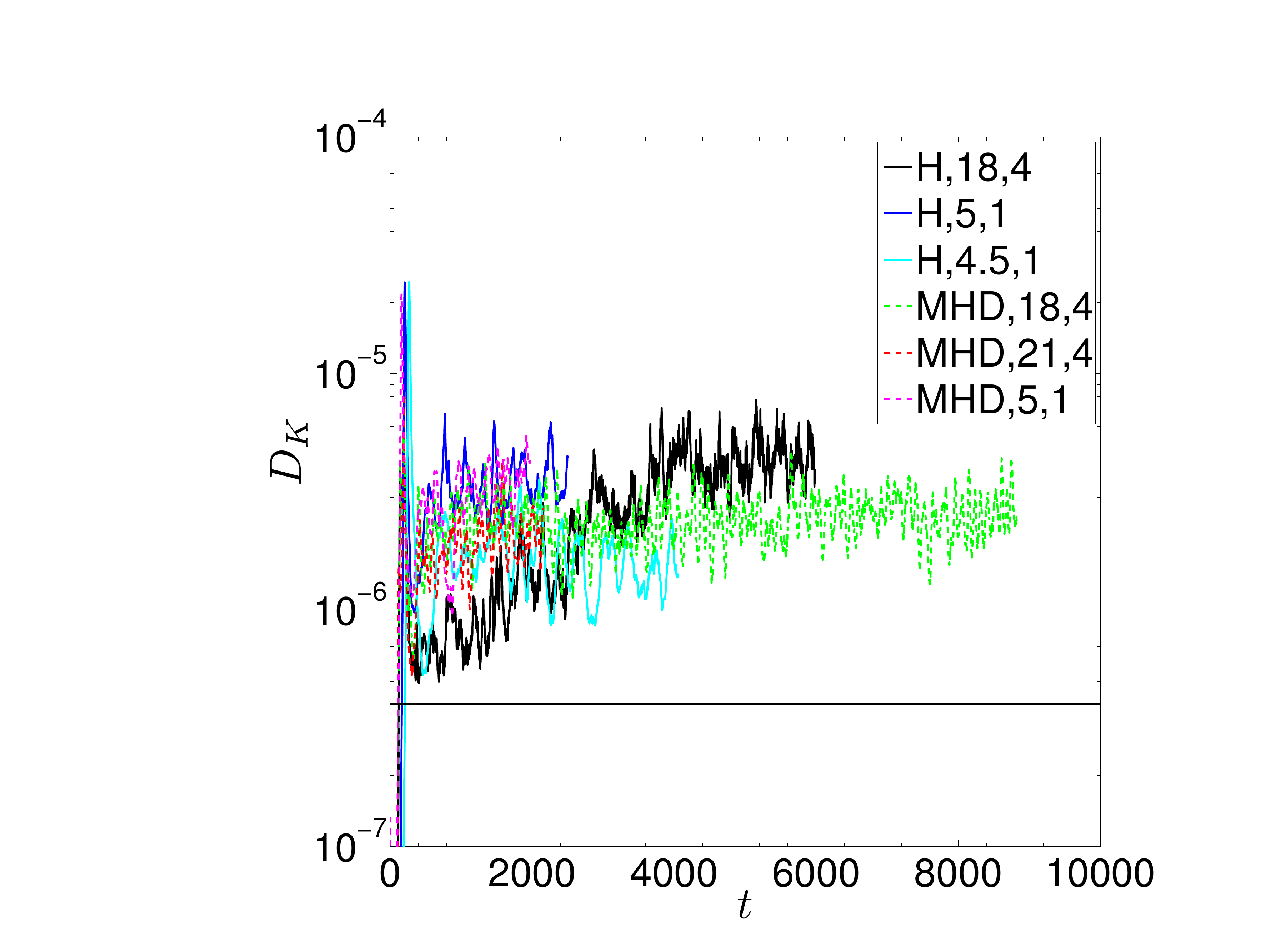}}
   \subfigure{\includegraphics[trim=6cm 0cm 7cm 1cm, clip=true,width=0.3\textwidth]{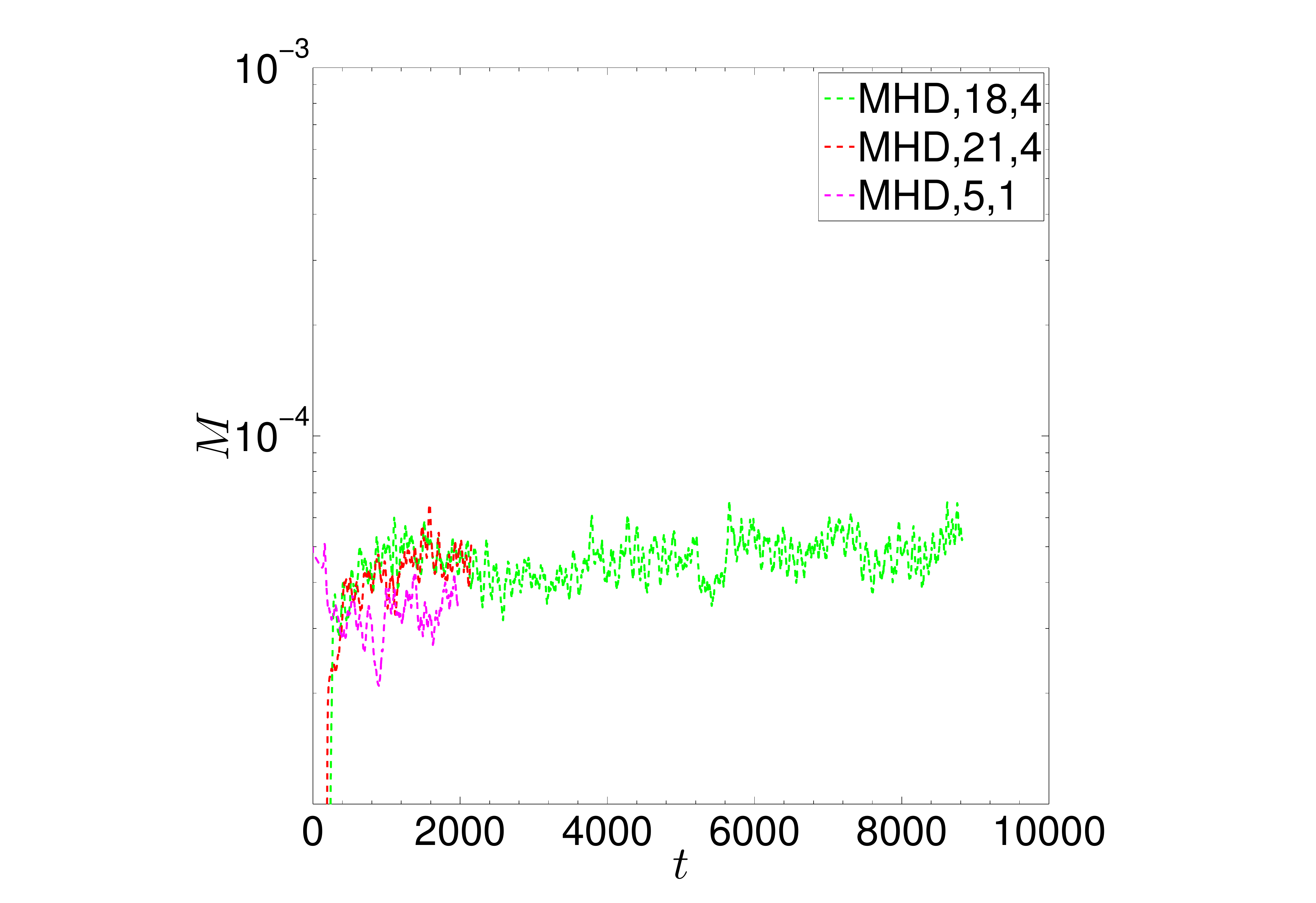}}
 \end{center}
  \caption{Temporal evolution of volume-averaged flow quantities from various hydrodynamical and MHD simulations with $\epsilon=0.1$, adopting either standard diffusion operators ($\alpha=1$) or hyperdiffusion (with $\alpha=4$). The label H,$18,4$ corresponds with a hydrodynamical simulation with $\alpha=4$ and $\nu_4=10^{-18}$, and similarly for other cases. Top: kinetic energy ($K=\langle \frac{1}{2}|\boldsymbol{u}|^2\rangle_V$). Second: RMS vertical velocity ($\bar{u}_z=\sqrt{\langle u_{z}^2 \rangle_V}$). Third: viscous dissipation rate ($D_K=-\nu_\alpha (-1)^{\alpha+1}\langle \boldsymbol{u}\cdot \nabla^{2\alpha} \boldsymbol{u} \rangle_V$), together with laminar viscous dissipation rate $D_\mathrm{lam}$ with $\nu=10^{-5}$ as the horizontal solid black line. Bottom: magnetic energy ($M=\langle \frac{1}{2}|\boldsymbol{B}|^2\rangle_V$).}
  \label{2}
\end{figure}

First, we describe several illustrative simulations that have $\epsilon=0.1$ with various diffusivities (values of $\nu_\alpha$) and diffusion operators (values of $\alpha$), with and without an initial magnetic field. In the purely hydrodynamical simulations, once the initial growth of the instability has saturated, the flow becomes organised into large-scale columnar vortices aligned with the rotation axis, as we illustrate in the top panel of Fig.~\ref{3}. This shows the vertical component of the vorticity ($\nabla\times\boldsymbol{u}$) at a chosen time in the simulation which is well after the saturation of the initial instability. The top panel of Fig.~\ref{2} shows the volume-averaged kinetic energy ($K=\langle \frac{1}{2}|\boldsymbol{u}|^2\rangle_V$, which represents the energy in both the waves and any large-scale vortices, if present, and $\langle \cdot \rangle_V = \frac{1}{L^3}\int_V \cdot \,\mathrm{d} V$) as a function of time, which continues to grow if there is only weak viscous dissipation on the large-scales, indicating that the large-scale vortices are continually driven. The black line shows a hydrodynamical simulation with $\alpha=4$ hyper-diffusion and $\nu_4=10^{-18}$ (H18), which continues to grow throughout the duration of this simulation, in contrast with the simulation represented by the light blue solid line, which has standard viscosity with $\alpha=1$ and $\nu_1=10^{-4.5}$ (H4.5), for which viscosity is able to resist this continual growth. An intermediate case with standard viscosity (H5) is plotted as the blue line. 

The presence of these vortices modifies the efficiency of wave excitation by the precessional instability, shown in the second panel of Fig.~\ref{2}, which plots the temporal evolution of the RMS vertical velocity ($\bar{u}_z=\sqrt{\langle u_{z}^2 \rangle_V}$; which primarily represents the instability-driven waves). In this case, the RMS vertical velocity is greater in the H18 simulation in comparison with the H5 simulation, presumably due to the additional instabilities of these large-scale vortices. The H4.5 simulation has weaker turbulent velocities because viscous damping is non-negligible on the energetically-dominant scales. In contrast to the mean velocities, the corresponding viscous dissipation rate appears to vary only weakly with the diffusivities (and diffusion operator) between these examples, with the H5 and H18 simulation having similar late-time values for the mean dissipation rate (H4.5 is about half as dissipative). The temporal evolution of the volume-averaged (viscous) dissipation rate ($D_K=-\nu_\alpha (-1)^{\alpha+1}\langle \boldsymbol{u}\cdot \nabla^{2\alpha} \boldsymbol{u} \rangle_V$) is shown in the third panel of Fig.~\ref{2}. For reference, we also plot the viscous dissipation that would result from the laminar precessional flow ($D_\mathrm{lam}$) for the case with $\nu=10^{-5}$ as a horizontal black line. In each case, we note that the turbulent dissipation far exceeds the laminar value.

\begin{figure}
  \begin{center}
    \subfigure{\includegraphics[trim=6cm 0cm 6cm 0cm, clip=true,width=0.36\textwidth]{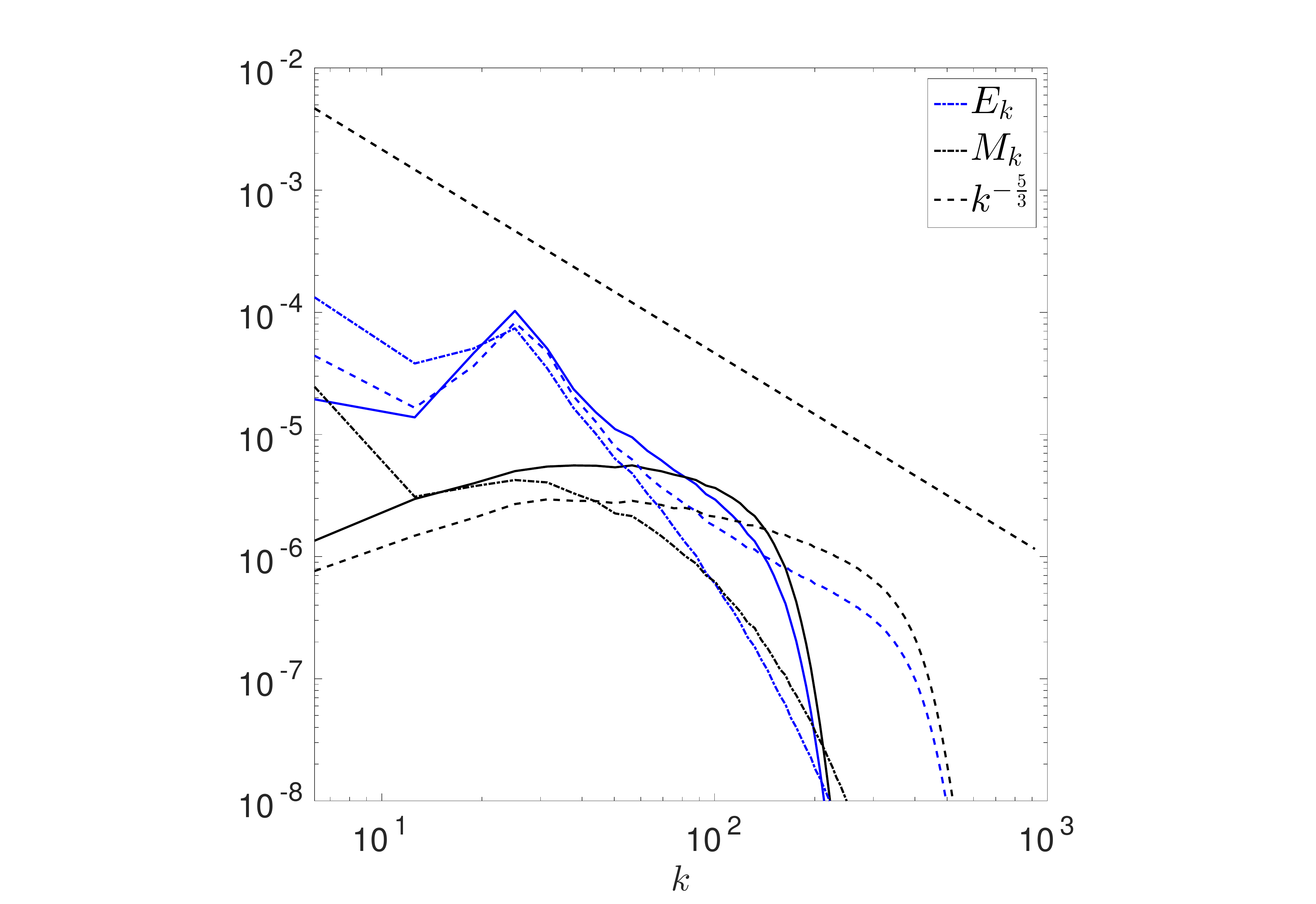}}
     \end{center}
  \caption{Kinetic (blue) and magnetic (black) energy (spherical-shell-averaged) spectra as a function of $k$ in three MHD simulations with $\epsilon=0.1$, with $\alpha=4,\nu_{4}=\eta_{4}=10^{-18}$ (solid), $\alpha=4,\nu_{4}=\eta_{4}=10^{-21}$ (dashed) and $\alpha=1,\nu_{1}=\eta_{1}=10^{-5}$ (dot-dashed). The kinetic energy at the energetically-dominant scales appears to depend only weakly on the diffusivities and diffusion operators in these cases. The Kolmogorov spectrum is plotted for reference as the black dashed slanted line.}
  \label{3a}
\end{figure}

In the simulations with an initially weak magnetic field, the initial instability behaves as it does in the hydrodynamical simulations. However, the magnetic field is subsequently amplified, and magnetic stresses either destroy or inhibit the formation of columnar vortices. This enables the turbulence to be sustained, with approximately constant mean kinetic energy, RMS vertical velocities and mean dissipation rates (only the viscous dissipation rate $D_K$ is plotted), once the system reaches a statistically steady state. The vertical vorticity in an example MHD simulation is plotted in the bottom panel of Fig.~\ref{3} at a chosen time snapshot, where this simulation has identical parameters with the top panel except for the presence of a magnetic field in the initial state. The presence of even a weak magnetic field significantly alters the properties of the turbulence driven by the precessional instability. However, the mean viscous dissipation rate for these examples is similar to the cases without a magnetic field (see the third panel of Fig.~\ref{2}). The growth of the magnetic field is illustrated in the bottom panel of Fig.~\ref{2}, where we plot the volume-averaged magnetic energy ($M=\langle \frac{1}{2}|\boldsymbol{B}|^2\rangle_V$). This demonstrates that the precessional instability-driven flow acts as a dynamo, which amplifies or maintains the magnetic energy. (We note that the magnetic field in an MHD simulation with $\nu_1=\eta_1=10^{-4.5}$ decays, so that its eventual evolution matches that of H4.5.)

To further probe the turbulent state, we illustrate the (spherical-shell-averaged) kinetic and magnetic energy spectra as a function of $k$ on Fig.~\ref{3a} for the MHD simulations just described. This shows the kinetic energy of the energetically-dominant scales to be approximately independent of the diffusivities and diffusion operator for these examples. In the hyperdiffusive simulations, there is a short inertial range that is roughly Kolmogorov-like, but this is less distinct in the example with standard viscosity and ohmic diffusivity. The magnetic energy is preferentially contained on small to intermediate-scales.

\begin{figure}
 \begin{center}
   \subfigure{\includegraphics[trim=6cm 0cm 7cm 1cm, clip=true,width=0.31\textwidth]{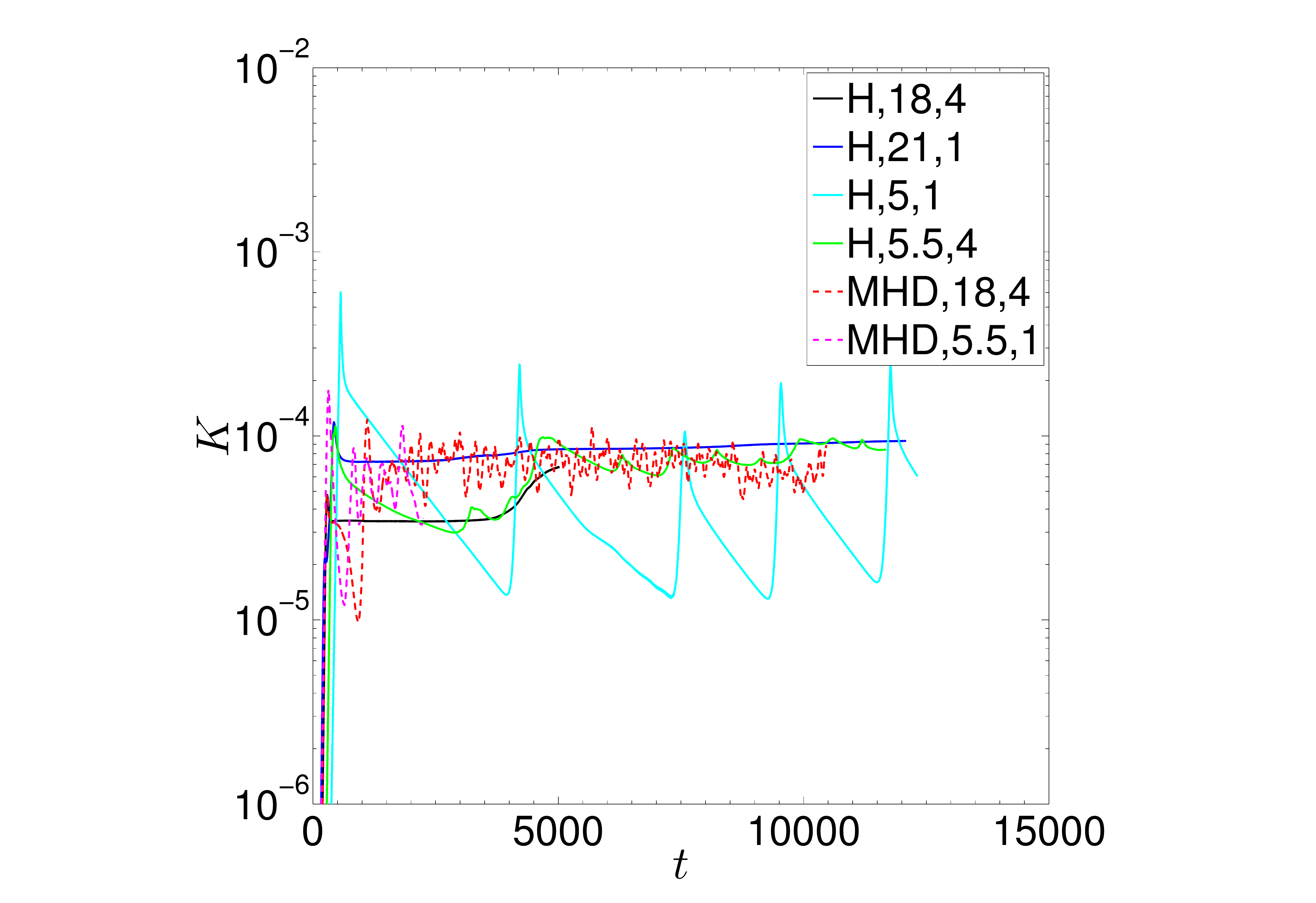}}
   \subfigure{\includegraphics[trim=8cm 0cm 2cm 3cm, clip=true,width=0.31\textwidth]{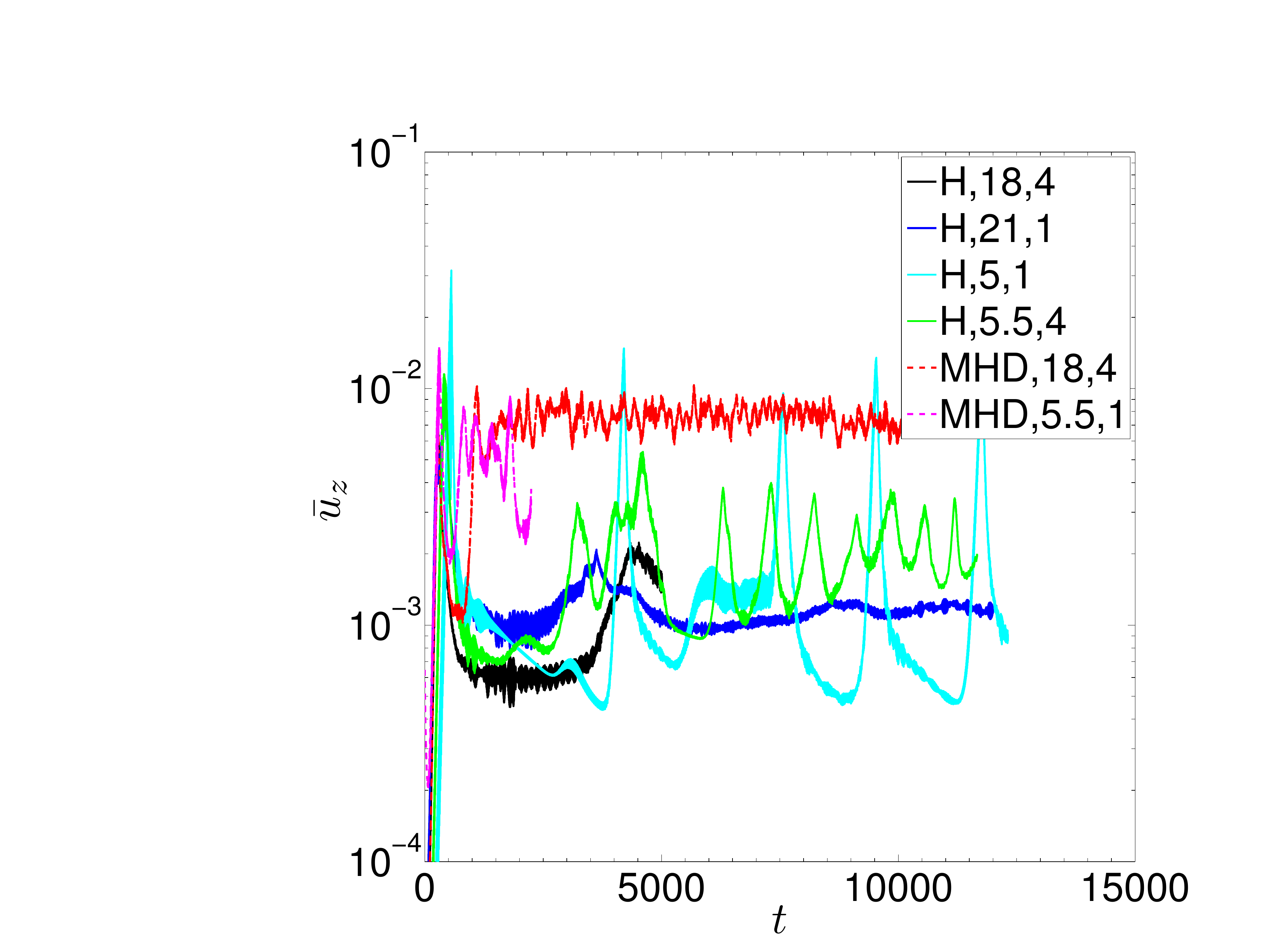}}
   \subfigure{\includegraphics[trim=7cm 0cm 2cm 3cm, clip=true,width=0.32\textwidth]{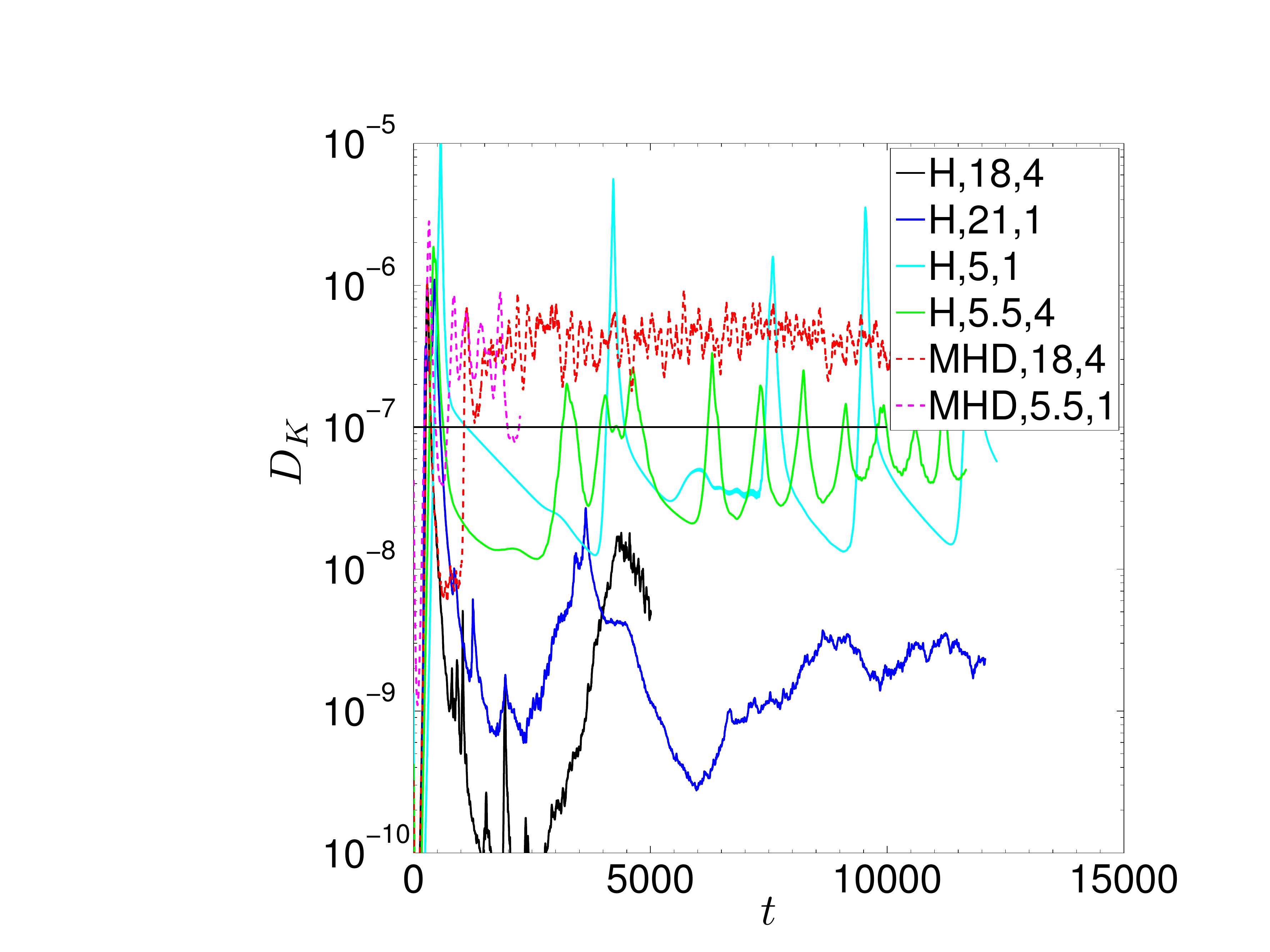}}
   \subfigure{\includegraphics[trim=6cm 0cm 7cm 1cm, clip=true,width=0.31\textwidth]{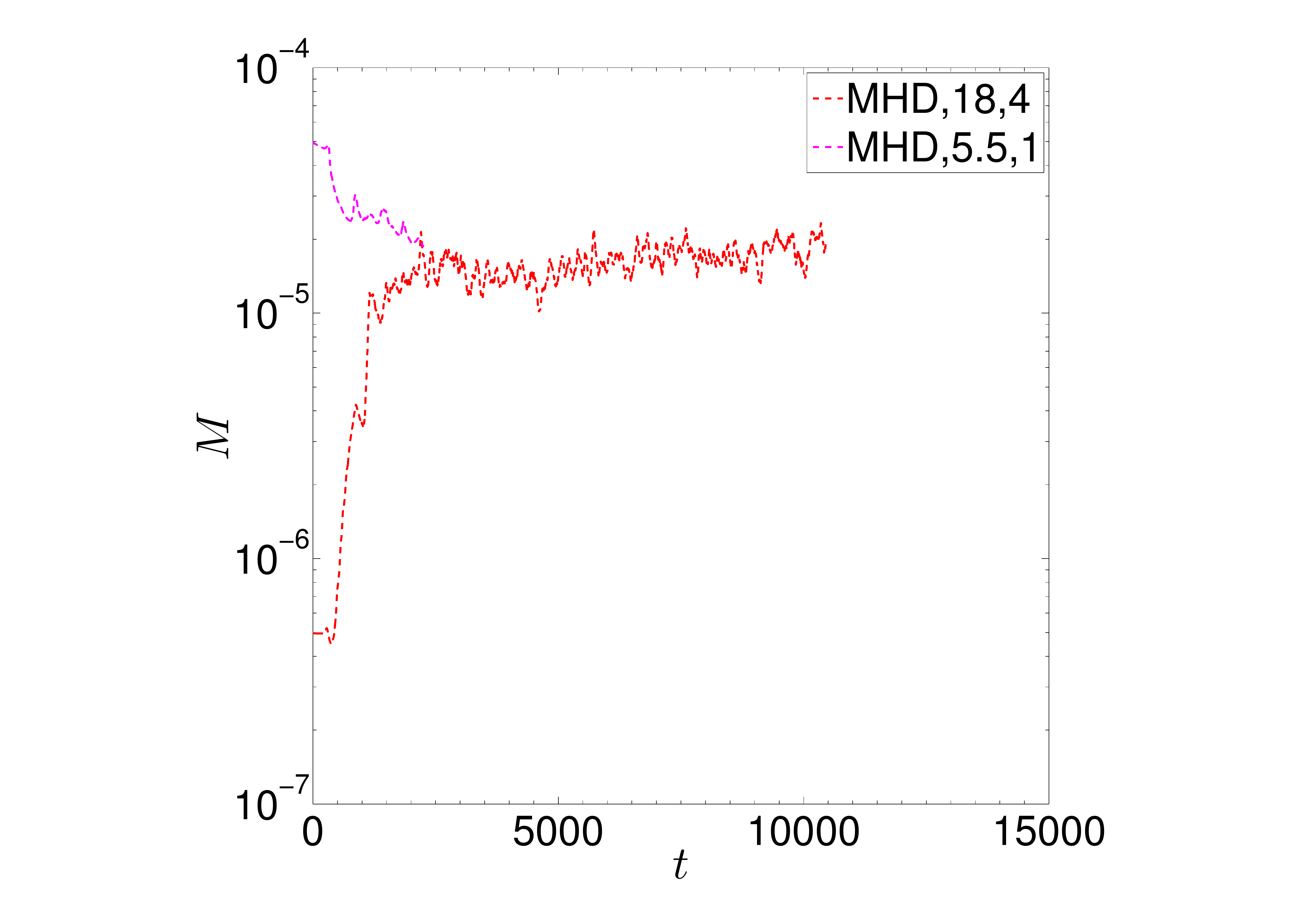}}
 \end{center}
  \caption{Same as Fig.~\ref{2}, but for a set of simulations with $\epsilon=0.05$.}
  \label{3b}
\end{figure}

In Fig.~\ref{3b} we plot the temporal evolution of volume-averaged quantities for a separate set of simulations that have $\epsilon=0.05$. As in Fig.~\ref{2}, this shows that even a weak magnetic field can alter the properties of the flow. The second and third panels illustrate that for $\epsilon=0.05$, the presence of columnar vortices in the hydrodynamical simulations acts to significantly reduce the turbulent velocities and dissipation rates, in comparison with the MHD cases, in which columnar vortex formation is prevented. The hydrodynamical simulation H5 exhibits strong bursty behaviour, associated with the formation and viscous damping of these vortices. In these examples, the presence of columnar vortices strongly inhibits the wave driving due to the precessional instability, and therefore reduces the efficiency of the turbulent dissipation. In the MHD simulations in Fig.~\ref{3b}, columnar vortices are not present, and the turbulence is sustained and statistically steady, exhibiting similar behaviour whether we adopt standard diffusion operators or hyperdiffusion. Once again, the flow acts as a dynamo, that generates and maintains a magnetic field.

\begin{figure}
 \begin{center}
  \subfigure{\includegraphics[trim=4cm 0cm 9cm 1cm, clip=true,width=0.34\textwidth]{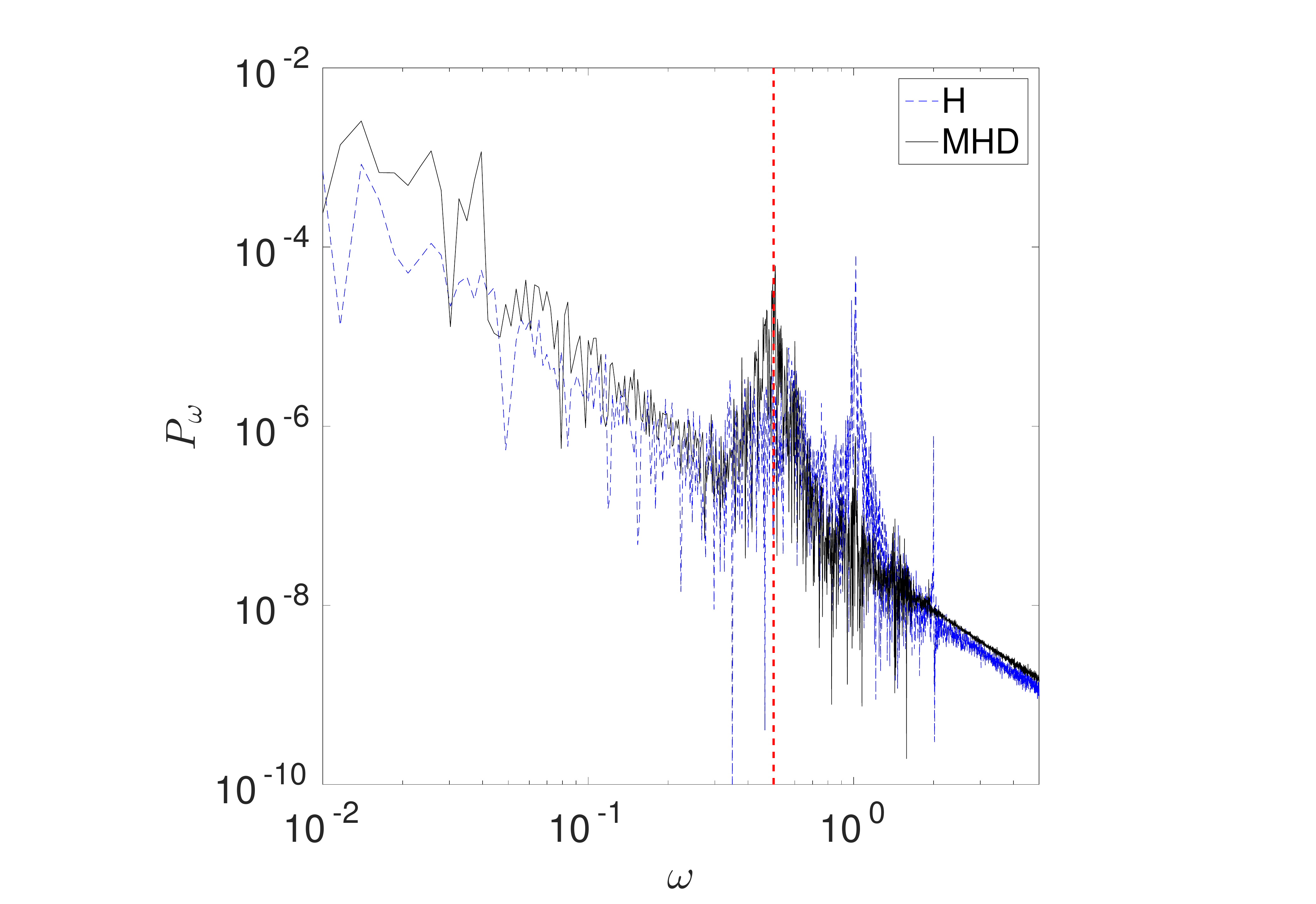}}
  \subfigure{\includegraphics[trim=4cm 0cm 9cm 1cm, clip=true,width=0.34\textwidth]{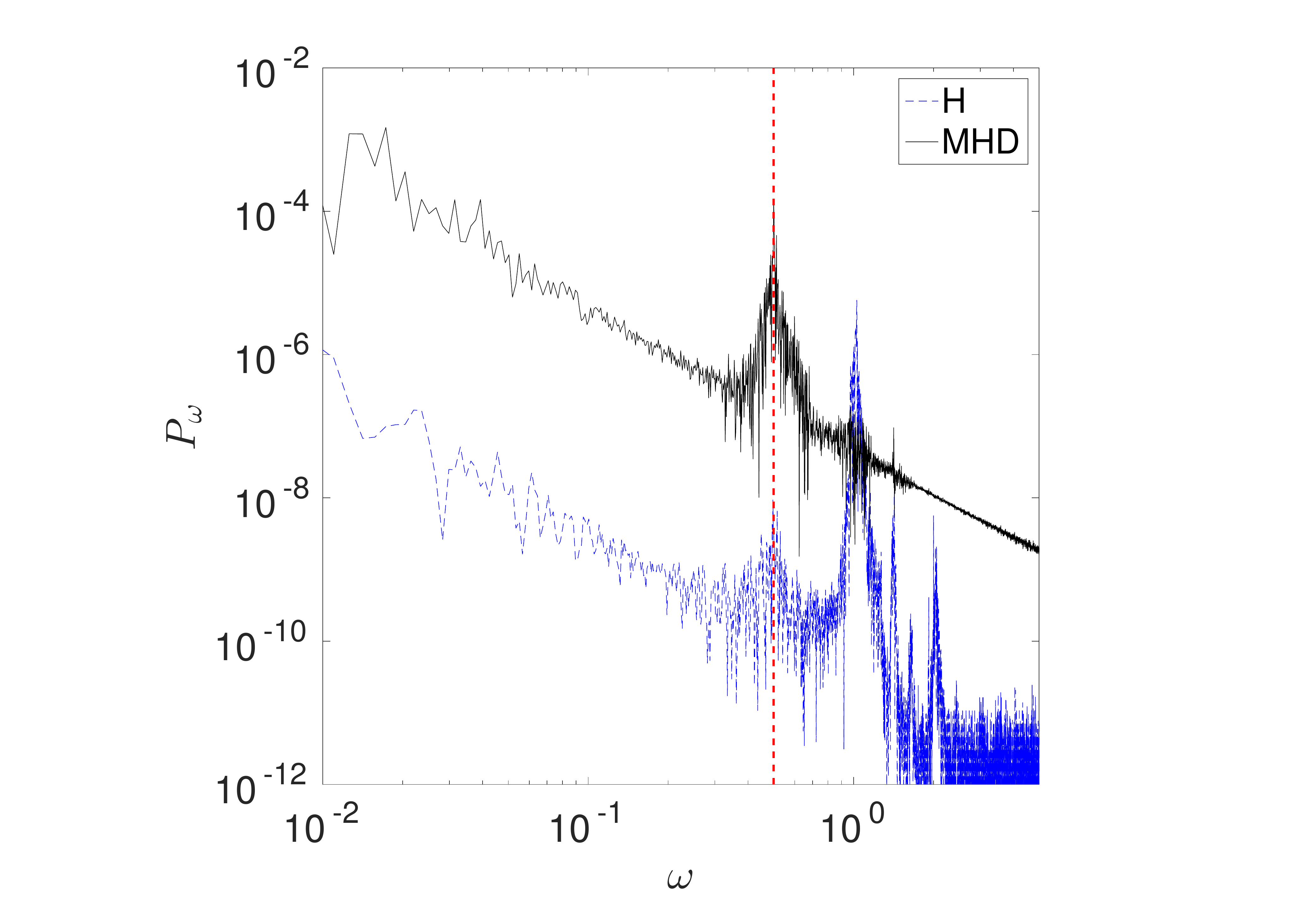}}
 \end{center}
  \caption{Temporal power spectrum $P_{\omega}=\frac{1}{2}|\hat{u}_z(\omega)|^2$ (where $\hat{u}_z (\omega)$ is the temporal Fourier transform of $\bar{u}_z$) for $\epsilon=0.1$ (top, using data from $t=250-1600$) and $\epsilon=0.05$ (bottom, using data from $t=1000-3000$), with $\alpha=4$ hyperdiffusion and $\nu_{4}=\eta_4=10^{-18}$ (except for the hydrodynamical simulation with $\epsilon=0.05$, for which $\nu_{4}=10^{-21}$). We have divided the frequencies in each panel by two since $\bar{u}_z$ oscillates with twice the frequency of the fluid at a test point. In both panels, we compare a hydrodynamic (blue dashed line) and an MHD simulation (solid black line). The predicted frequency of the linear precessional instability ($|\omega|=\frac{1}{2}$) is illustrated by the vertical red dashed lines.}
  \label{3c}
\end{figure}

To further analyse these simulations, and probe the differences between the hydrodynamic and MHD simulations, we plot the temporal power spectrum of the RMS vertical velocity ($\bar{u}_z$) in Fig.~\ref{3c} for two sets of simulations with $\epsilon=0.1$ (top panel) and $\epsilon=0.05$ (bottom panel) -- we have divided the frequencies by two since $\bar{u}_z$ oscillates with twice the frequency of the fluid at a test point. In both panels, the MHD simulations exhibit a clear peak at $\omega=\frac{1}{2}$, which corresponds with the dominant frequency of the linear precessional instability, as expected. However, the dominant peak in the hydrodynamical simulations occurs instead at $\omega=1$, which corresponds with direct forcing at the precession frequency. This is due to forcing of the large-scale vortices, and only in some cases does this lead to a sustained energy injection into the flow (see Fig.~\ref{2}). Frequencies with $\omega=\frac{1}{2}$ are suppressed (relative to the MHD cases) due to the presence of columnar vortices that inhibit the precessional instability. In both cases, there is also a concentration of power in very low frequencies. Fig.~\ref{3c} further highlights the differences in nonlinear evolution with and without a magnetic field.

To summarise these results, we have found that the precessional instability leads to turbulence and to enhanced viscous dissipation over that of the laminar precessional flow. In hydrodynamical simulations, the flow becomes organised into columnar vortices, which modify the resulting wave excitation and strongly inhibit it if $\epsilon=0.05$. The columnar vortices in this Cartesian model will presumably correspond with zonal flows in a more realistic global model -- indeed, this is the case for the related elliptical instability \citep{BL2013,Barker2015b}. In the presence of a weak magnetic field, columnar vortices are no longer produced, and the turbulence is sustained and statistically steady. In many ways, the outcome of the precessional instability that we have just described is very similar to the related elliptical instability \citep{BL2013,BL2014}.

\section{Synthesis of results}
\label{synthesis}

\begin{figure*}
  \begin{center}
    \subfigure{\includegraphics[trim=5cm 0cm 6cm 0cm, clip=true,width=0.36\textwidth]{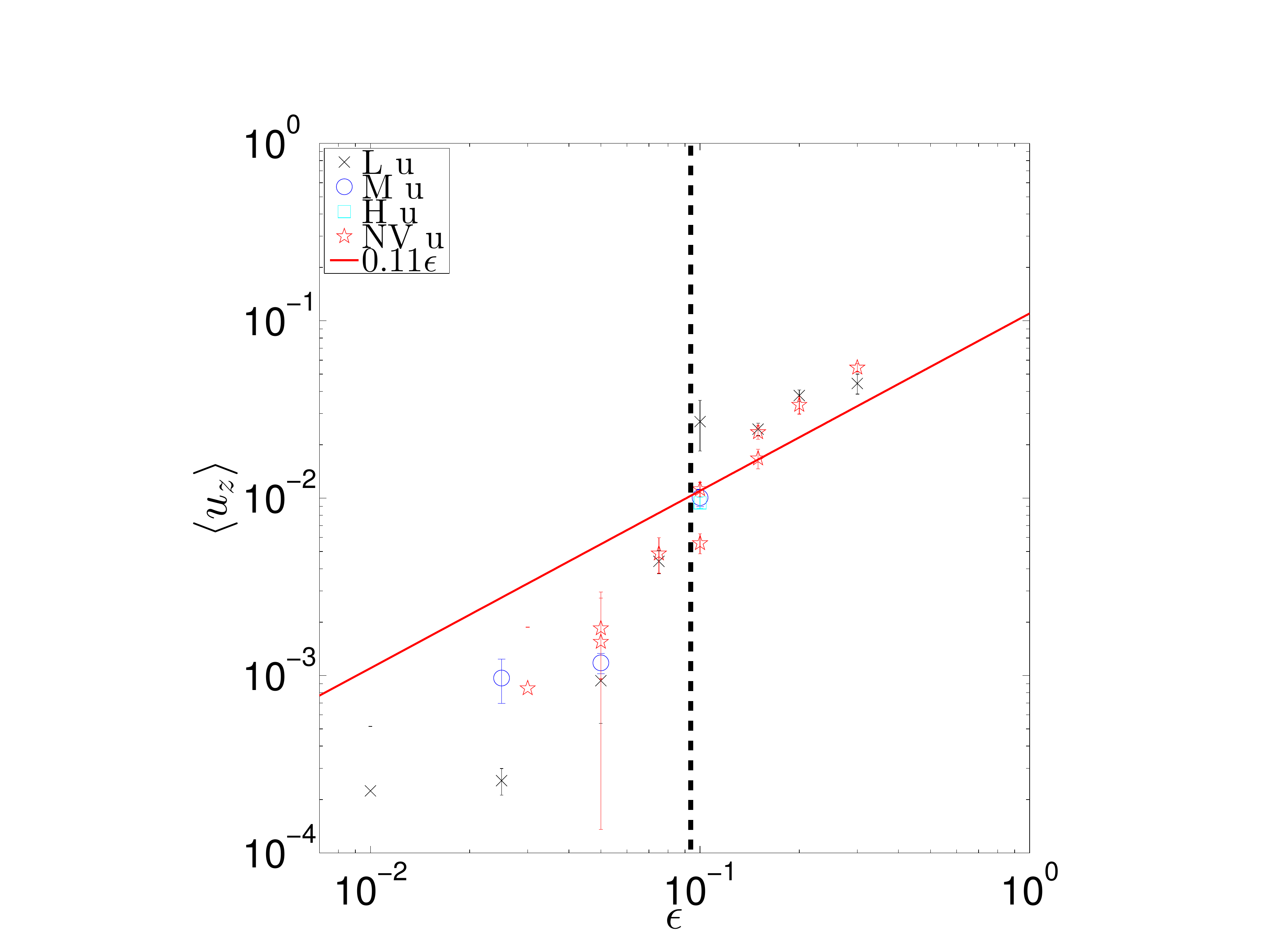}}
    \subfigure{\includegraphics[trim=5cm 0cm 6cm 0cm, clip=true,width=0.4\textwidth]{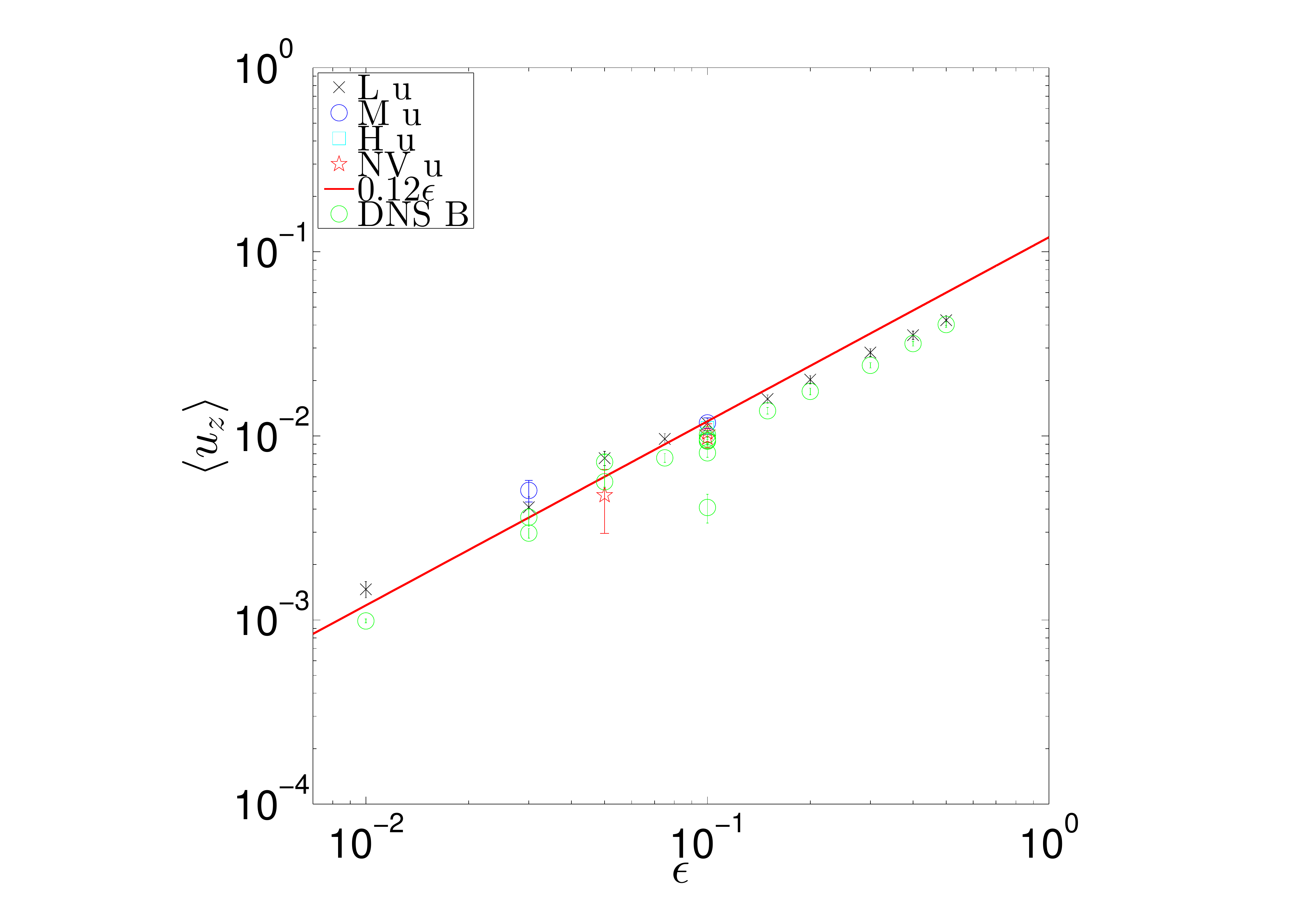}}
    \subfigure{\includegraphics[trim=5cm 0cm 6cm 0cm, clip=true,width=0.36\textwidth]{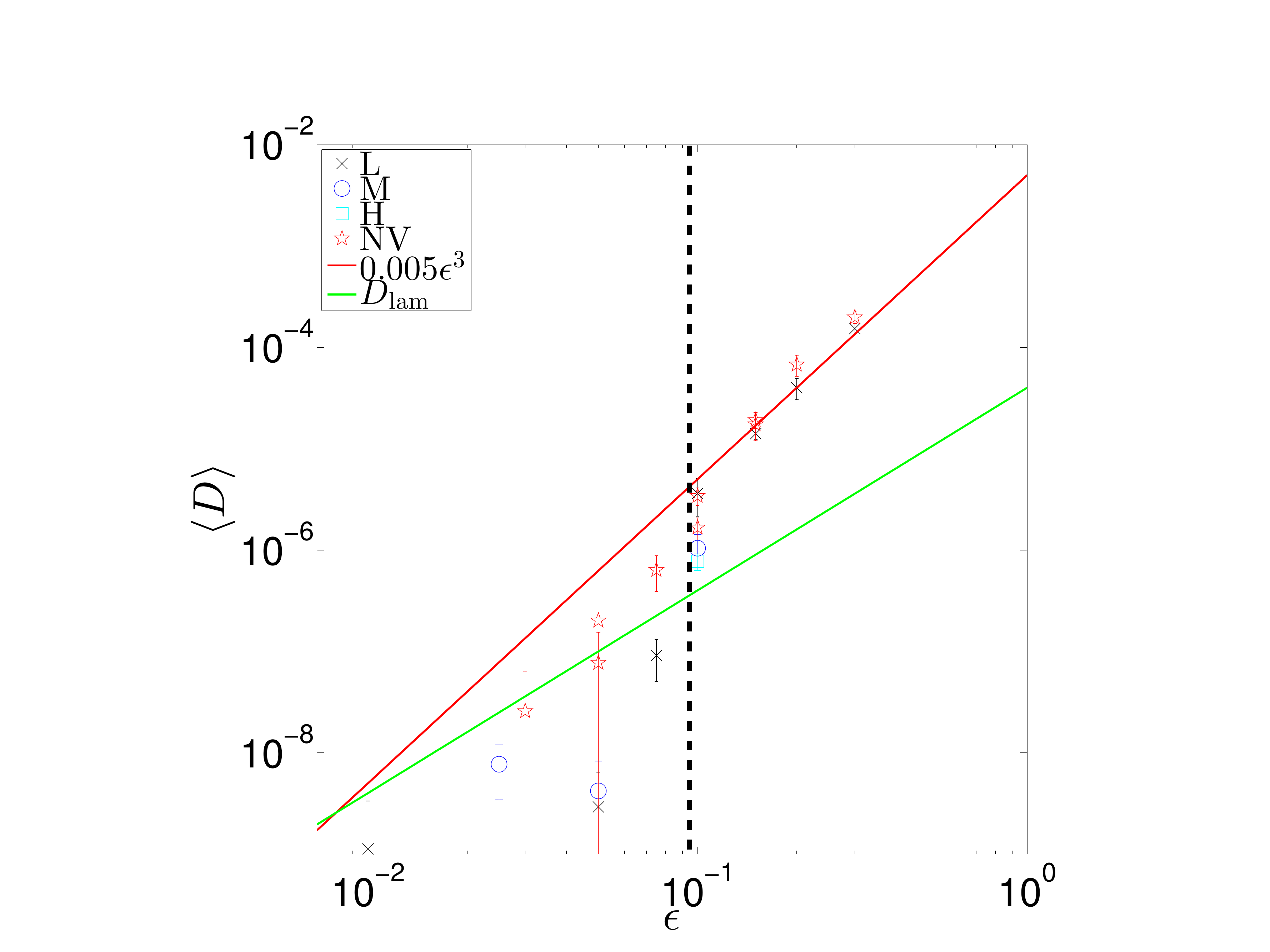}}
    \subfigure{\includegraphics[trim=5cm 0cm 6cm 0cm, clip=true,width=0.4\textwidth]{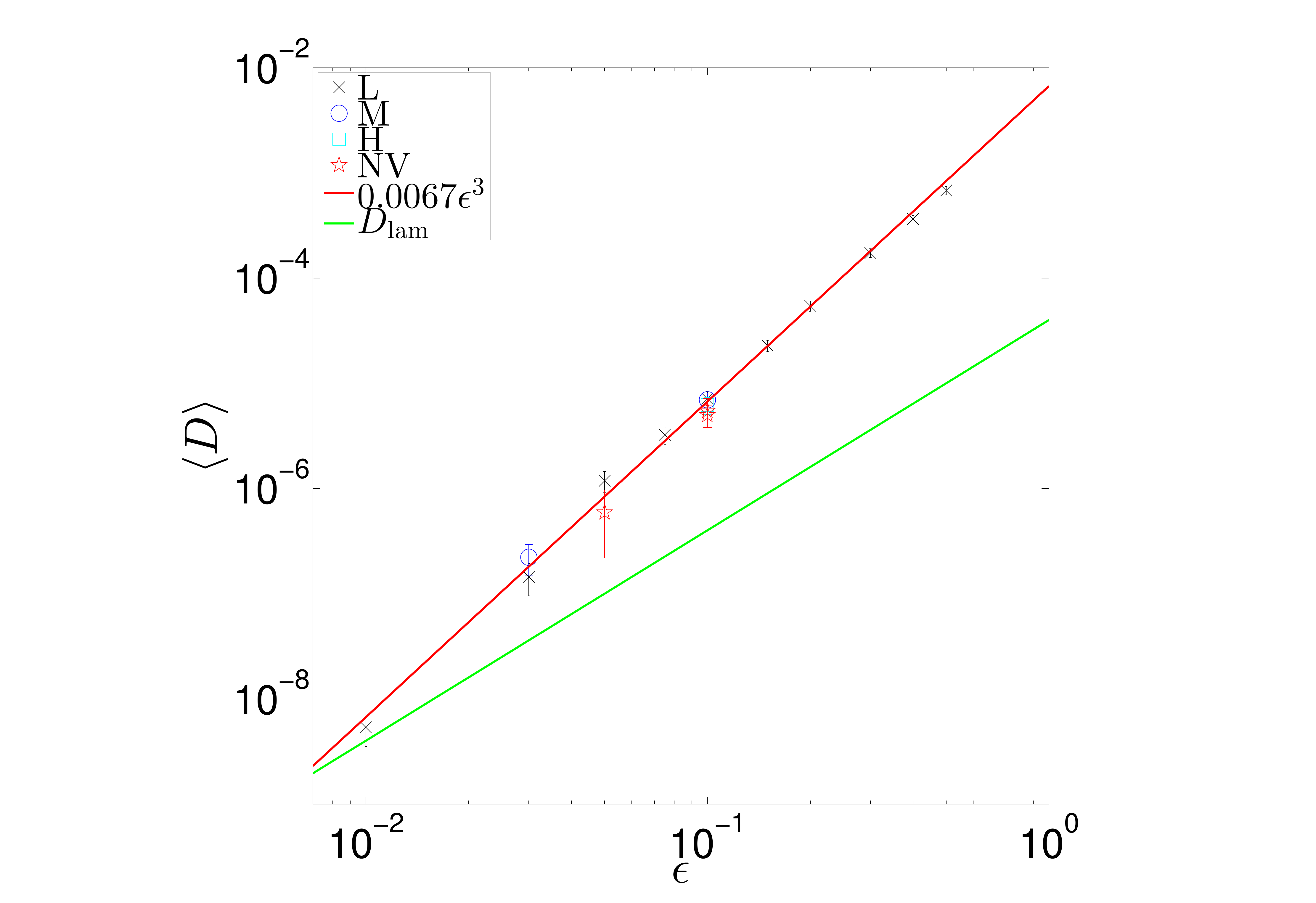}}
    \end{center}
  \caption{Synthesis of simulation results, plotting time-averaged RMS turbulent velocities (top panels) and (viscous plus ohmic) dissipation rates (bottom panels) as a function of $\epsilon$ for the precessional instability. Results from hydrodynamical simulations are presented in the left panels and those from MHD simulations with an initially weak magnetic field in the right panels. The MHD results support the scalings of \S~\ref{turb} with $\langle u_z\rangle\approx 0.12\epsilon$ and $\langle D\rangle\approx0.0067\epsilon^3$, approximately independent of the diffusivities. We have also plotted the laminar viscous dissipation ($D_\mathrm{lam}$) as a green solid line, based on assuming $\nu=10^{-5}$, for reference.  Simulations with several different diffusivities and diffusion operators have also been plotted according to the legend (and Table~\ref{table}), where NV denotes normal viscosity and L, M and H are the lowest effective resolution to highest effective resolution hyperdiffusive runs. In the hydrodynamical simulations, columnar vortices reduce the turbulent intensity for $\epsilon \lesssim 0.1$, evidenced by the drop-off in the scalings with $\epsilon$ -- this regime transition is indicated by the vertical black dashed lines. The green circles in the top right panel show the RMS magnetic field $\langle B\rangle=\sqrt{\langle 2M\rangle}$.}
  \label{4}
\end{figure*}
The primary purpose of this paper is to determine the astrophysical importance of the precessional instability. In order to extrapolate to the astrophysical regime in which $\epsilon\ll1$, we have run a suite of simulations, as listed in Table~\ref{table}, to study the variation of the mean turbulent velocity and mean turbulent dissipation as a function of $\epsilon$, as well as the diffusivities and diffusion operators (as far as we are able to vary these). These simulations were run at modest resolutions to permit a wider parameter survey, and also because they are run for a long duration in comparison with a typical turbulent turnover time.

Fig.~\ref{4} is the main result of this paper, and shows the variation in the time-averaged RMS vertical velocity ($\langle u_z\rangle\equiv \langle \bar{u}_z\rangle$, where angled brackets without subscripts denote a time average) and total (viscous and ohmic) dissipation rate ($\langle D\rangle\equiv \langle D_K+D_M\rangle$). The left panels show the results of hydrodynamical simulations (in which $D_M=0$) whereas the right panels show the results of MHD simulations with an initially weak magnetic field. The predicted scalings from \S~\ref{turb} are indicated by the red lines, where $\chi$ and $C$ have been fitted (by eye) to best describe the data. Error bars represent RMS fluctuations, and the viscous dissipation of the laminar precessional flow ($D_\mathrm{lam}$) with $\nu=10^{-5}$ is indicated using the green line for illustration. Simulations with normal viscosity (and ohmic diffusivity) are represented using red stars (NV in the legend), whereas those with hyperdiffusion are represented using black crosses (L), blue circles (M) and light blue squares (H), respectively. The label L refers to the lowest effective resolution using hyperdiffusivities, i.e., the largest diffusivity, or smallest value of $-\log_{10} \nu_{4}$, whereas H refers to the highest effective resolution and M to the intermediate case.

The MHD simulations provide strong support for the scalings predicted in \S~\ref{turb} for $\epsilon\in[0.01,0.5]$, which we would expect for a statistically steady turbulent cascade. In particular, the RMS vertical velocities are well described by\footnote{A linear velocity scaling was also found by \cite{WuRoberts2008}, who performed simulations using a similar Cartesian model, but with rigid stress-free upper and lower boundaries.} $\langle u_z\rangle\approx C\epsilon$, with $C\approx 0.12$. The mean turbulent (viscous and ohmic) dissipation is consistent with $\langle D\rangle\approx \chi \epsilon^3$, with $\chi\approx0.0067$, and is typically significantly larger than the viscous dissipation of the laminar precessional flow (for $\nu=10^{-5}$). In most cases, variation of the diffusivities (and diffusion operators), appears to modify the turbulent velocities and dissipation rates only weakly, at least as far as these simulations are able to probe. Since $\sigma_{\mathrm{max}}/|\epsilon |$ is decreasing function of $\epsilon$, we would expect $\langle u_z\rangle$ to be slightly smaller than this estimate for $\epsilon\gtrsim 0.1$, which is consistent with what we observe.

On the other hand, the hydrodynamical simulations exhibit a regime transition for $\epsilon\approx 0.1$ (crudely indicated by the vertical dashed lines), above which the results are roughly consistent with an $\epsilon^3$ scaling (with a similar coefficient to the MHD simulations), but below which there is a significant departure in the scaling. The latter is due to the inhibition of wave driving by the precessional instability when the flow is organised into large-scale columnar vortices. This reduces the resulting turbulent velocities and dissipation rates for $\epsilon\lesssim 0.1$, but this drop-off is less pronounced with normal viscosity and strongest with hyperdiffusion. Indeed, the simulations with normal viscosity approximately support an $\epsilon^3$ scaling for the dissipation rate for $\epsilon\in[0.03,0.3]$, with $\langle D\rangle \approx 0.005\epsilon^3$. Note that $\langle D\rangle$ is approximately independent of the diffusivities for $\epsilon\gtrsim 0.1$, but strongly depends on the diffusivities for $\epsilon\lesssim 0.1$.

\begin{figure}
  \begin{center}
    \subfigure{\includegraphics[trim=6cm 0cm 6cm 0cm, clip=true,width=0.35\textwidth]{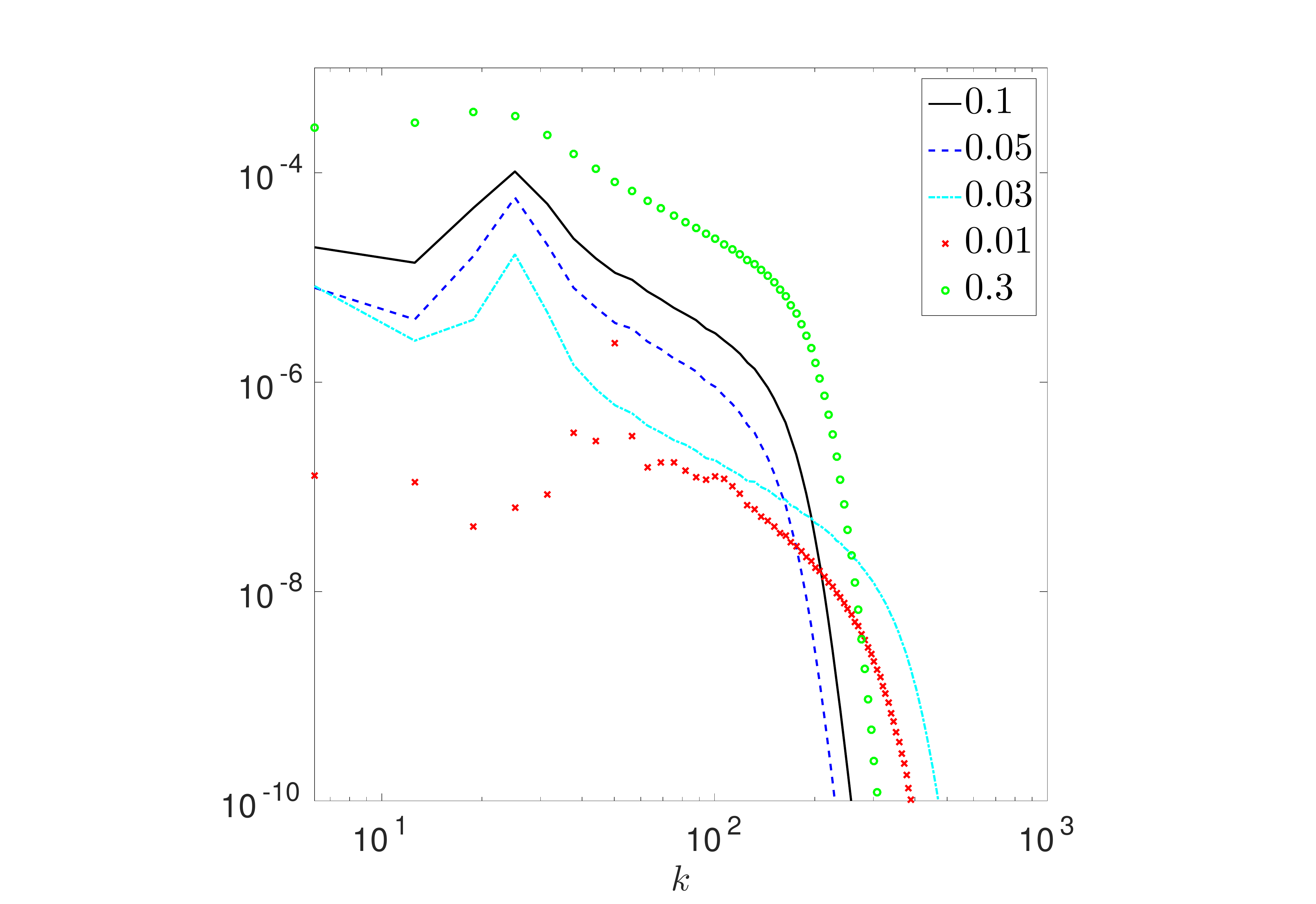}}
     \end{center}
  \caption{Kinetic energy (spherical-shell-averaged) spectra as a function of $k$ for a set of MHD simulations with various $\epsilon$, as indicated in the legend. These adopted $\alpha=4,\nu_{4}=\eta_{4}=10^{-18}$ (for $\epsilon=0.05,0.1,0.3$) and $\alpha=4,\nu_{4}=\eta_{4}=10^{-21}$ (for $\epsilon=0.01,0.03$). This indicates that the energetically-dominant wavenumber may depend weakly on $\epsilon$.}
  \label{4a}
\end{figure}

We have also analysed the kinetic energy spectra for several MHD simulations to determine whether the energetically-dominant scales exhibit any dependence on $\epsilon$. This is plotted in Fig.~\ref{4a}, in which we observe that the dominant wavenumber is: $k\approx 18$ for $\epsilon=0.3$, $k\approx 25$ for $\epsilon\in[0.03,0.1]$, and $k\approx 55$ for $\epsilon=0.01$. This suggests a weak dependence of the energetically-dominant wavelength $\lambda\sim k^{-1}$ on $\epsilon$. 

Such a trend might be expected because instability only occurs in frequency bands of width $\Delta \omega=O(\epsilon\Omega)$ around exact resonance. Since there are only a finite number of global modes with $\lambda\sim L$ (the size of the box, or the planetary radius), the probability that a box-scale (or planetary-scale) mode is excited becomes very small as $\epsilon\Omega \rightarrow 0$. However, resonances will always be found on small-enough scales (in the absence of viscosity) as long as $\epsilon\Omega\ne 0$. The number of modes with a given minimum wavelength $\lambda$ scales as $(\frac{\lambda}{L})^{-3}$. Therefore a mode with a minimum wavelength $\lambda$ has a reasonable probability to be excited if $\frac{\Delta\omega}{2\Omega} \left(\frac{\lambda}{L}\right)^{-3}\gtrsim O(1)$. This suggests $\lambda \propto \epsilon^{\frac{1}{3}}$, i.e., $k\propto\epsilon^{-\frac{1}{3}}$, which is consistent with the trend that we have observed\footnote{I thank Jeremy Goodman for pointing out this argument for the related elliptical instability (see \citealt{Barker2015b}).}. 

If we otherwise apply the arguments of \S~\ref{turb}, this would suggest $\langle D\rangle \propto \epsilon^{\frac{11}{3}}$ and $\langle u_z\rangle \propto \epsilon^{\frac{4}{3}}$, instead of $\epsilon^3$ and $\epsilon$, respectively, i.e., $\chi\propto \epsilon^{\frac{2}{3}}$ and $C\propto \epsilon^{\frac{1}{3}}$, instead of being constants. This would make the instability less efficient at small $\epsilon$ than suggested by the estimates of \S~\ref{turb}, and would predict a stronger dependence of the spin-orbit evolutionary timescale on orbital period. A crude way of estimating the implications of this is to take $\chi\approx 10^{-2}\left(\epsilon/0.1\right)^{\frac{2}{3}}$ in Eq.~\ref{taupred}. This would modify our prediction for the period out to which this instability could be important (i.e.~$\tau_{\psi}\sim 1$ Gyr) to approximately 10 d rather than 18 d. However, Fig.~\ref{4} indicates that $\langle D\rangle$ (and $\langle u_z\rangle$) are remarkably consistent with an $\epsilon^3$ (and $\epsilon$) scaling, and does not appear consistent with an $\epsilon^{\frac{11}{3}}$ (and $\epsilon^{\frac{4}{3}}$) scaling over the simulated range of parameters.

In summary, we have shown that in the presence of a magnetic field, the turbulence driven by the precessional instability is well described by the scalings of \S~\ref{turb} over the range $\epsilon\in[0.01,0.5]$ that we can probe numerically. This suggests that the tidal evolutionary timescales in \S~\ref{turb} may be estimated by assuming $\chi\approx 0.01$, which represents an approximate upper bound on the turbulent dissipation efficiency based on these simulations. However, as we have discussed, it is possible these scalings would no longer hold for even smaller $\epsilon$. Further calculations that probe more deeply into the small $\epsilon$ regime would be worthwhile to determine whether this is the case.

\subsection{Dynamo driven by precession}
\label{dynamo}

\begin{figure}
  \begin{center}
     \subfigure{\includegraphics[trim=6cm 0cm 6cm 0cm, clip=true,width=0.35\textwidth]{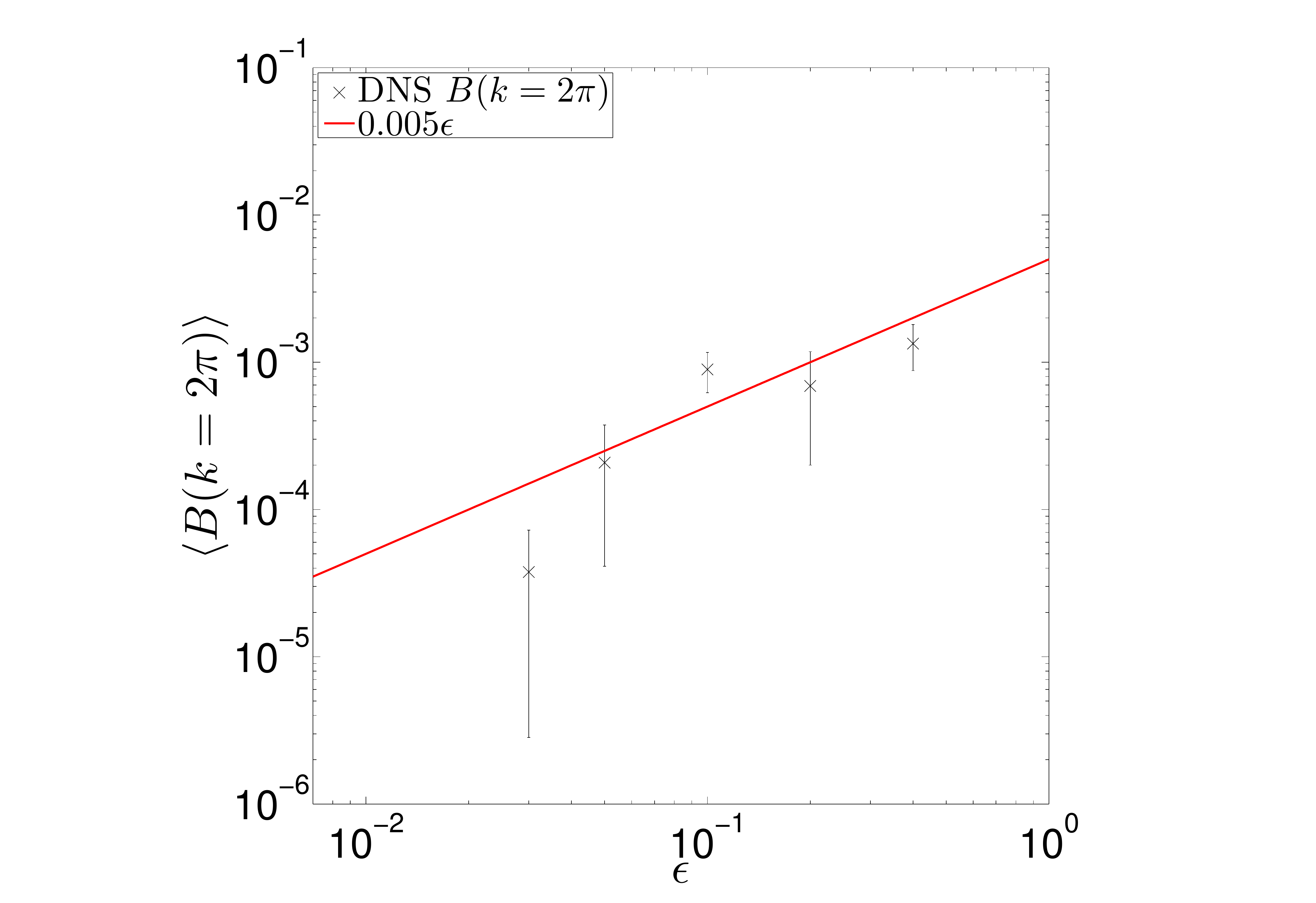}}
     \end{center}
  \caption{Mean box-scale magnetic field ($\langle B(k=2\pi)\rangle$) on the largest scales, with $k=2\pi$ as a function of $\epsilon$. For reference, we have plotted the line $0.005\epsilon$, which is roughly consistent with the data for $\epsilon\gtrsim 0.05$. Whether or not this scaling continues to be valid for small $\epsilon$, this demonstrates that the precessional instability can generate box-scale magnetic fields, and motivates further study to determine whether they may be important in driving a system-scale dynamo.}
  \label{5}
\end{figure}

Our secondary aim is to study the ability of the precessional instability to drive a dynamo. We now turn to discuss this aspect of our simulation results. We have already shown in Figs.~\ref{2} and \ref{3b} that the precessional instability can amplify a small initial magnetic field and subsequently maintain the magnetic energy. In the top right panel of Fig.~\ref{4}, we have plotted the time-averaged RMS magnetic field ($\langle B\rangle \equiv\sqrt{\langle 2M\rangle}$) as the green circles. This is found to follow a similar scaling as the turbulent velocities, with the RMS magnetic field satisfying $\langle B\rangle\approx C_B\epsilon$), with $C_B\approx0.09$ (so that $M<K$, but not strongly so). As we have shown in Fig.~\ref{3a}, the magnetic energy is preferentially on small-scales. However, there is non-negligible magnetic energy in the box-scale components of the magnetic field (with $k=2\pi$). 

In Fig.~\ref{5}, we plot the RMS magnetic field in the $k=2\pi$ component as a function of $\epsilon$ for a separate set of simulations that were initialised with white-noise perturbations to the velocity and magnetic field (for all wavenumbers $\frac{k}{2\pi}\in[2,21]$ with amplitude $10^{-5}$). In these simulations, there is no magnetic energy in the $k=2\pi$ component initially, but this is subsequently generated by the turbulence. These results are approximately consistent with $B(k=2\pi)\approx 0.005 \epsilon$, but there appears to be a drop-off for small $\epsilon$, perhaps indicating that there is a threshold $\epsilon$ for system-scale dynamo action. 

There are significant uncertainties in applying these scalings to predict the magnetic fields strengths that could be generated in real bodies, partly because our diffusivities are much larger than in reality, and partly because we have only considered identical viscosity and magnetic diffusivity (magnetic Prandtl numbers equal to unity). Nevertheless, our simulations suggest that the precessional instability is able to drive a system-scale dynamo for these parameters. This may play a role in generating (at least in part) the magnetic fields of precessing short-period planets. Further work is required to verify whether this in indeed the case, using more realistic global simulations, particularly those with small magnetic Prandtl numbers (i.e.~ohmic diffusivities that are much larger than the viscosity).

\section{Conclusions}

We have studied the hydrodynamical instabilities of the flows induced by axial precession in giant planets (and stars). Our primary aim was to determine the importance of the resulting turbulence in driving evolution of the spin--orbit angle (obliquity) in giant planets (and stars). To do this, we adopted an idealised Cartesian model that can be thought to represent a small-patch in the convective interior of a giant planet (or star). This model allowed us to simulate the nonlinear outcome of the instabilities of precession-driven flows, with and without a weak magnetic field.

We found that the turbulent dissipation resulting from the precessional instability (in the presence of a weak magnetic field) is consistent with $\langle D\rangle \approx \chi M_p\Omega_p^3R_p^2$, where $\chi\approx 10^{-2}$ can be taken to provide an approximate upper bound, and $M_p, R_p$ and $\Omega_p$ are the planetary mass, radius and precession frequency, respectively. (In the limit of very weak precession, when $\Omega_p\ll\Omega$, where $\Omega$ is the planetary spin frequency, it is possible -- indeed, likely -- that $\chi$ would exhibit a weak dependence on $\Omega_p/\Omega$, but this is not suggested by our existing numerical results for $\langle D\rangle$ in the range $0.01 \lesssim \Omega_p/\Omega\lesssim 0.5$.)

This dissipation is sufficiently strong that it could play an important role in driving tidal evolution of the spin--orbit angle for hot Jupiters with orbital periods shorter than about 10--18 days (or perhaps longer if the planet can remain inflated). In isolation, this instability would drive evolution towards 0 or 180 degrees, depending on the initial spin--orbit angle. But in the presence of other tidal mechanisms, the ultimate evolution would be towards alignment. Other mechanisms that may be important include inertial waves excited in the presence of a core (e.g.~\citealt{Gio2004,GoodmanLackner2009,Ogilvie2013,FBBO2014}), dissipation in the core itself \citep{Remus2012,Storch2014}, as well as the elliptical instability (e.g.~\citealt{Barker2015b}). The precessional instability is likely to provide an additional source of tidal dissipation in planets with spin--orbit misalignments. However, this mechanism appears unlikely to be important in driving evolution of the spin--orbit angle in stars with planetary-mass companions, or in close binary systems, since the precession in both cases is generally much slower. This would lead to relatively weak turbulent dissipation and correspondingly long evolutionary timescales.

Our results suggest that photometric observations of transiting planets would be unlikely to observe axial precession (due to transit depth variations, e.g.~\citealt{CarterWinn2010a,CarterWinn2010b}) in giant planets with orbital periods shorter than about 10--18 days. This is because this instability, acting in isolation, would predict such planets to have undergone significant tidal evolution towards alignment or anti-alignment, where precession would no longer occur. However, if any planets initially had spin--orbit angles close to 90 degrees, the predicted evolutionary timescales would be much longer, so the axial precession of this subset of planets may be observable. If planetary axial precession is observed through future photometric studies, it would place important constraints on the mechanisms of tidal dissipation in giant planets and on their tidal evolutionary histories.

In our hydrodynamical simulations, the instability led to turbulence and the formation of columnar vortices, which produced burstiness in the dissipation and significantly reduced the turbulent intensity. These columnar vortices are commonly found in turbulence subjected to rapid rotation, and would be expected to correspond with zonal flows in a global model \citep{Barker2015b}. In the presence of even a weak magnetic field, the properties of the flow are significantly altered, and the formation of these vortices is inhibited. This permits sustained turbulence that is statistically steady, exhibiting enhanced dissipation over the case without a magnetic field in simulations with small $\epsilon$. This is very similar to what is found for the related elliptical instability \citep{BL2013,BL2014}. 

We have also shown that the precessional instability can drive a dynamo, which may be able to produce system-scale magnetic fields. This suggests that the magnetic fields of short-period gaseous planets may be generated, at least in part, by this mechanism. Based on these indicative results, we would recommend further work to study dynamos driven by precession (or the related elliptical instability) in more realistic global simulations (continuations of e.g.~\citealt{Tilgner2005,WuRoberts2009,ErnstHullerman2013}).

The calculations presented in this paper should be regarded as a first step towards understanding the turbulent damping of axial precession in giant planets (and stars). In particular, it is not clear whether the base flow driven by precession is well represented by a simple, steady Poincar\'{e}-like flow (Eq.~\ref{basicflow}). Indeed, in a fluid body with a \textit{rigid} outer boundary in the shape of a triaxial ellipsoid, the equivalent internal laminar flow can exhibit non-steady behaviour \citep{NoirCebron2013}. In addition, for our simple estimates, we have assumed that the vertical shearing strain (which drives the instabilities that we study) is equal to $\Omega_p$, but it is possible that this is an underestimate. It is important to determine whether or not this is the case, because the dissipation appears to scale as the cube of this quantity. Nevertheless, it is hoped that the calculations presented here may shed light on the turbulence driven by the precessional instability, and its potential effects on the spin--orbit evolution of giant planets (or stars).

Further work is required to analyse precession-driven flows in more realistic (linear and nonlinear) global models of (perhaps differentially) rotating giant planets or stars. It is possible that geometrical effects may then modify the laminar precession-driven flow and its corresponding turbulent dissipation rates. In addition, the presence of an inner core may lead to somewhat enhanced dissipation due to the excitation of inertial waves from the core (e.g.~\citealt{HollerbachKerswell1995}). Finally, the interaction of the flows driven by this instability with turbulent convective motions should be considered, as should its coexistence with the elliptical instability.

\section*{Acknowledgements}
I would like to thank Yufeng Lin for useful discussions and for generously commenting on an earlier draft of this paper, and Jeremy Goodman, the referee, for recommending the inclusion of Fig.~\ref{3c}. This work was supported by the Leverhulme Trust and Isaac Newton Trust through the award of an Early Career Fellowship.

\appendix

\section{MHD equations in the local model of precession}
\label{MHD}
The extension of the local model to study perturbations to a background precessional flow (cf.~Eq.~\ref{perturbationeqns}) to non-ideal magnetohydrodynamics (using units such that $\sqrt{\mu_0 \rho}=1$) is given by 
\begin{eqnarray}
\nonumber
&&D\boldsymbol{u} = -\nabla p -2 \boldsymbol{e}_z\times \boldsymbol{u}-\mathrm{A}\boldsymbol{u}-2\boldsymbol{\epsilon}(t)\times\boldsymbol{u}\\
&& \hspace{1cm} +\boldsymbol{B}\cdot \nabla\boldsymbol{B}+D^\nu_\alpha (\boldsymbol{u}), \\
&& D\boldsymbol{B}=\boldsymbol{B}\cdot \nabla\boldsymbol{u} +\mathrm{A}\boldsymbol{B} + D^\eta_\alpha (\boldsymbol{B}), \\
&& \nabla\cdot\boldsymbol{u}=0, \\
&& \nabla\cdot\boldsymbol{B}=0,
\end{eqnarray} 
where
\begin{eqnarray}
&& D\equiv\partial_t +\boldsymbol{u}\cdot \nabla +\mathrm{A}\boldsymbol{x}\cdot\nabla, \\
&& D^\nu_\alpha (\boldsymbol{u}) \equiv\nu_\alpha (-1)^{\alpha+1}\nabla^{2\alpha}\boldsymbol{u}, \\
&& D^\eta_\alpha (\boldsymbol{B}) \equiv\eta_\alpha (-1)^{\alpha+1}\nabla^{2\alpha}\boldsymbol{B}.
\end{eqnarray} 
As in \S~\ref{spectral}, we solve these equations using a basis of time-dependent wavevectors (also for $\boldsymbol{B}$), so that their solutions can be computed using a Fourier spectral code. The kinetic ($K=\langle \frac{1}{2}|\boldsymbol{u}|^2\rangle_V$), magnetic ($M=\langle \frac{1}{2}|\boldsymbol{B}|^2\rangle_V$) and total energy equations ($E=K+M$) are
\begin{eqnarray}
&& \partial_t K = -\langle \boldsymbol{u}\mathrm{A}\boldsymbol{u}\rangle_V + \langle \boldsymbol{u}\cdot \left(\boldsymbol{B}\cdot \nabla\boldsymbol{B}\right)\rangle_V-D_K, \\
&& \partial_t M = \langle \boldsymbol{B}\mathrm{A}\boldsymbol{B}\rangle_V - \langle \boldsymbol{u}\cdot \left(\boldsymbol{B}\cdot \nabla\boldsymbol{B}\right)\rangle_V-D_M, \\
&& \partial_t E = -\langle \boldsymbol{u}\mathrm{A}\boldsymbol{u}\rangle_V + \langle \boldsymbol{B}\mathrm{A}\boldsymbol{B}\rangle_V - D_K-D_M,
\end{eqnarray}
where $D_K=-\nu_\alpha (-1)^{\alpha+1}\langle \boldsymbol{u}\cdot \nabla^{2\alpha} \boldsymbol{u} \rangle_V$, $D_M=-\eta_\alpha (-1)^{\alpha+1}\langle \boldsymbol{B}\cdot \nabla^{2\alpha} \boldsymbol{B} \rangle_V$, and $D=D_K+D_M$ is the mean total (viscous and ohmic) dissipation rate. We have defined $\langle \cdot \rangle_V = \frac{1}{L^3}\int_V \cdot \, \mathrm{d}V$. If $\alpha=1$, the diffusion operators reduce to standard viscosity and ohmic diffusion, but $\alpha>1$ ``hyperdiffusion" is adopted in some simulations, which reduces diffusion on the large scales. This is done in order to determine its effects, and because it enables a greater range of $\epsilon$ to be simulated.

In most of the simulations, we initialise the magnetic field to be large-scale and to have zero net-flux, of the form
\begin{eqnarray}
\boldsymbol{B}(\boldsymbol{x},t=0)=B_0 \sin \left(\frac{2\pi x}{L}\right) \boldsymbol{e}_z,
\end{eqnarray}
where we typically take $B_0=10^{-3}$ initially. However, in some simulations we instead initialise the magnetic field with white noise with amplitude $10^{-5}$ for all wavenumbers $k\in[2,21]$.

\section{Table of simulations}
\label{TableSims}
A table of simulations is presented in Table~\ref{table}.
\begin{table}
\begin{tabular}{cccccc}
 $\epsilon$ & $\boldsymbol{B}_0$ & $\alpha$ & $-\log_{10}\nu_\alpha$ & $N$ & Label/Comments \\
  \hline
 0.01 & 0 & 4 & 18 & 128 & L \\
 0.025 & 0 & 4 & 18 & 128 & L \\
 0.025 & 0 & 4 & 21 & 256 & M \\
 0.05 & 0 & 4 & 18 & 128 & L \\
 0.05 & 0 & 4 & 21 & 256 & M \\
 0.05 & 0 & 1 & 5.5 & 256 & NV \\
 0.075 & 0 & 4 & 18 & 128 & L \\
 0.075 & 0 & 1 & 5 & 128 & NV \\
 0.1 & 0 & 4 & 16 & 128 & L \\
 0.1 & 0 & 4 & 18 & 128 & M \\
 0.1 & 0 & 4 & 19 & 128 & H \\
 0.1 & 0 & 1 & 4.5 & 128 & NV \\
 0.1 & 0 & 1 & 5 & 256 & NV \\
 0.15 & 0 & 4 & 18 & 128 & L \\
 0.15 & 0 & 1 & 4.5 & 256 & NV \\
 0.15 & 0 & 1 & 5 & 256 & NV \\
 0.2 & 0 & 4 & 18 & 128 & NV \\
 0.2 & 0 & 1 & 4.5 & 256 & NV \\
 0.3 & 0 & 4 & 17 & 256 & NV \\
 0.3 & 0 & 1 & 4.5 & 256 & NV \\
 \hline
 0.01 & $10^{-3}$ & 4 & 21 & 256 & L \\
 0.03 & $10^{-3}$ & 4 & 18 & 128 & L \\
 0.03 & $10^{-3}$ & 4 & 21 & 256 & M \\
 0.03 & $10^{-5}$ & 4 & 21 & 256 & White Noise $\boldsymbol{B}$ \\
 0.05 & $10^{-3}$ & 4 & 18 & 128 & L \\
 0.05 & $10^{-5}$ & 4 & 18 & 128 & White Noise $\boldsymbol{B}$ \\
 0.05 & $10^{-2}$ & 1 & 5.5 & 128 & NV \\
 0.075 & $10^{-3}$ & 4 & 18 & 128 & L \\
 0.1 & $10^{-3}$ & 4 & 17 & 128 & L \\
 0.1 & $10^{-3}$ & 4 & 18 & 128 & M \\
 0.1 & $10^{-5}$ & 4 & 18 & 128 & White Noise $\boldsymbol{B}$ \\
 0.1 & $10^{-3}$ & 4 & 21 & 256 & H \\
 0.1 & $10^{-2}$ & 1 & 5 & 256 & NV \\
 0.1 & $10^{-2}$ & 1 & 5.5 & 256 & NV \\
 0.15 & $10^{-3}$ & 4 & 18 & 128 & L \\
 0.2 & $10^{-3}$ & 4 & 18 & 128 & L \\
 0.2 & $10^{-5}$ & 4 & 18 & 128 & White Noise $\boldsymbol{B}$ \\
 0.3 & $10^{-3}$ & 4 & 18 & 128 & L \\
 0.4 & $10^{-3}$ & 4 & 18 & 128 & L \\
 0.4 & $10^{-5}$ & 4 & 18 & 128 & White Noise $\boldsymbol{B}$ \\
 0.5 & $10^{-3}$ & 4 & 18 & 128 & L \\
\end{tabular}
\caption{Table of simulations performed for this work. In all MHD cases $\nu_\alpha=\eta_\alpha$. $N$ is the number of Fourier modes in each dimension.}
\label{table}
\end{table}

\bibliography{tid}
\bibliographystyle{mn2e}
\label{lastpage}
\end{document}